\documentclass[12pt]{article}

\AtBeginDocument{
  \addtocontents{toc}{\normalsize}
}
\pdfoutput=1

\usepackage{amsmath,amssymb,amsfonts,amsthm,amscd,mathrsfs}
\usepackage{xcolor}
\definecolor{darkblue}{rgb}{0.1,0.1,.7}
\usepackage[colorlinks, linkcolor=darkblue, citecolor=darkblue, urlcolor=darkblue, linktocpage]{hyperref} 
\usepackage[]{version}
\usepackage[]{graphicx}
\usepackage[]{latexsym}
\usepackage{geometry}
\usepackage[all,cmtip]{xy}
\usepackage[margin=10pt,font=small,labelfont=bf]{caption}
\usepackage{ifthen}
\usepackage{tikz}
\usepackage{array,setspace,mathrsfs,amsfonts,yfonts,dsfont,bbm,colonequals}
\usepackage{dsfont}
\usepackage{cite}
\usepackage{xspace}

\numberwithin{equation}{section}

\newcommand{\reef}[1]{(\ref{#1})}
\newcommand{\be}{\begin{equation}}
\newcommand{\ee}{\end{equation}}
\newcommand{\bea}{\begin{eqnarray}}
\newcommand{\eea}{\end{eqnarray}}
\newcommand{\ba}{\begin{equation}\begin{aligned}}
\newcommand{\ea}{\end{aligned}\end{equation}}

\newcommand{\ud}{\mathrm d}

\newcommand{\Df}{\Delta_\phi}

\newcommand{\De}{\Delta}

\newcommand{\Db}{\Delta_b}

\begin{document}

\vspace*{-.6in} \thispagestyle{empty}
\begin{flushright}
LPTENS/18/05
\end{flushright}
\vspace{1cm} {\Large
\begin{center}
{\bf The Analytic Functional Bootstrap I:\\1D CFTs and 2D S-Matrices}\\
\end{center}}
\vspace{1cm}
\begin{center}
{\bf Dalimil Maz\'a\v{c}$^{a,b}$, Miguel F.~Paulos$^{c}$}\\[1cm] 
{
\small
$^{a}$ {\em C. N. Yang Institute for Theoretical Physics, SUNY, Stony Brook, NY 11794-3840, USA}\\
$^{b}$ {\em Simons Center for Geometry and Physics, SUNY, Stony Brook, NY 11794-3636, USA}\\
$^{c}$ {\em Laboratoire de Physique Th\'eorique de l'\'Ecole Normale Sup\'erieure\\ PSL University, CNRS, Sorbonne Universit\'es, UPMC Univ. Paris 06\\ 24 rue Lhomond, 75231 Paris Cedex 05, France
}
\normalsize
}
\\
\end{center}

\vspace{4mm}

\begin{abstract}
We study a general class of functionals providing an analytic handle on the conformal bootstrap equations in one dimension. We explicitly identify the extremal functionals, corresponding to theories saturating conformal bootstrap bounds, in two regimes. The first corresponds to functionals that annihilate the generalized free fermion spectrum. In this case, we analytically find both OPE and gap maximization functionals proving the extremality of the generalized free fermion solution to crossing. Secondly, we consider a scaling limit where all conformal dimensions become large, equivalent to the large $AdS$ radius limit of gapped theories in $AdS_2$. In this regime we demonstrate analytically that optimal bounds on OPE coefficients lead to extremal solutions to crossing arising from integrable field theories placed in large $AdS_2$. In the process, we uncover a close connection between asymptotic extremal functionals and S-matrices of integrable field theories in flat space and explain how 2D S-matrix bootstrap results can be derived from the 1D conformal bootstrap equations. These points illustrate that our formalism is capable of capturing non-trivial solutions of CFT crossing.

\end{abstract}
\vspace{2in}


\newpage

{
\setlength{\parskip}{0.05in}
\tableofcontents
\renewcommand{\baselinestretch}{1.0}\normalsize
}


\setlength{\parskip}{0.1in}
\newpage

\section{Introduction}
The wonderful fact that highly non-trivial bounds on general conformal field theories (CFTs) can arise from mere unitarity and crossing symmetry of four-point correlators was first established in the ground-breaking work of \cite{Rattazzi:2008pe}. Since then the conformal bootstrap philosophy \cite{Poland:2016chs}  has proved to be a powerful principle, capable of producing qualitative and quantitative insights into the physics of strongly-coupled CFTs, such as in the context of critical phenomena and holographic dualities.\footnote{See \cite{Ferrara:1973yt,Polyakov:1974gs,Mack:1975jr} for early formulations of the conformal bootstrap and \cite{Rychkov:2016iqz,Simmons-Duffin:2016gjk} for pedagogical reviews of the modern developments.}

The conformal bootstrap is the program of extracting information from CFT crossing equations. Based on the physical picture of \cite{Alday:2007mf}, an exact analysis of these equations at large spin has led to a number of important results, starting with the pioneering works of \cite{Komargodski:2012ek,Fitzpatrick:2012yx}, finding an elegant systematization in \cite{Alday:2015eya,Alday:2015ewa,Alday:2016njk} and culminating in Caron-Huot's OPE inversion formula \cite{Caron-Huot2017b}. However, the task of determining optimal bounds on the CFT data of operators with low dimension and spin has remained mostly in the realm of numerical algorithms (implemented for example in \cite{Paulos:2014vya,Simmons-Duffin:2015qma}).

The initial steps towards an analytic understanding of the bootstrap bounds were undertaken in \cite{Mazac:2016qev}, where an optimal bound on the gap in 1D CFTs was shown to be saturated by a free fermion. The main goal of the present work is to extract the essential features of that construction and use them to start a systematic exploration of analytic bootstrap bounds. Most importantly, we will demonstrate that this approach is general enough to describe interacting solutions to CFT crossing.

In the first paper of this series, we will focus on the comparatively simpler but still highly non-trivial case of one-dimensional CFTs. We would like to emphasize that even here, our understanding of unitary solutions to crossing symmetry is severely limited. This is partly because the 1D bootstrap equations do not admit an expansion which is under control simultaneously in both channels, as opposed to the situation in higher D. At the same time, in a sense all local operators in 1D are scalars and thus understanding 1D crossing illuminates precisely the aspects of higher-D crossing not addressed by the light-cone bootstrap. Our approach can therefore be thought of as complementary to the existing work on the analytic bootstrap. However, we will make the case in subsequent work \cite{Mazac:2018ycv} that functional-based methods have the potential to subsume and generalize existing approaches, be that the light-cone bootstrap or the Polyakov-style bootstrap of \cite{Polyakov:1974gs,Gopakumar2017,Gopakumar2017a}.

Let us explain our approach in more detail. A conformal bootstrap equation arises by matching different OPE expansions of correlation functions. It is parametrized
by a set of cross-ratios and thus implies a continuously infinite set of constraints on the CFT data. These equations take the form of sum rules where each primary operator in the spectrum contributes additively. Hence the equations naturally live in a certain infinite-dimensional vector space of functions of the cross-ratios. It is useful to parametrize the constraints by the elements of the dual space, which are usually called functionals. The simplest example of a functional is the evaluation of the bootstrap equation at any point in the cross-ratio space where both OPE expansions can be trusted. Bounds on the CFT data arise from functionals with specific positivity properties, with optimal bounds corresponding to distinguished elements of the dual space, called \emph{extremal functionals} \cite{ElShowk:2012hu}. In principle a bound can only be optimal if there is a solution to crossing which saturates it, and therefore extremal functionals typically contain enough information to reconstruct this solution uniquely. For example, the operators exchanged in the OPE of the extremal solution must correspond to zeros of the extremal functional. To analytically obtain a conformal bootstrap bound means to analytically construct the appropriate extremal functional. Our approach in this series of works is to understand the physics and the construction of these functionals, and how they determine the properties of CFTs.

The first analytic examples of extremal functionals were constructed in \cite{Mazac:2016qev}. The specific functionals prove that the optimal upper bound on the scaling dimension of the leading operator above identity in the OPE of identical primaries $\phi$ in unitary 1D CFTs is $\Delta_{\textrm{gap}} = 2\Delta_\phi + 1$. The unique solution to crossing saturating this bound is the four-point function of the generalized free fermion,\footnote{Equivalently, this theory describes boundary correlators of a free massive fermion in $AdS_2$.} the leading non-identity operator in the OPE being the bilinear operator $\phi\!\!\overleftrightarrow{\partial}\!\!\phi$. The functionals were constructed for $\Delta_\phi$ a positive integer or half-integer and take the form of contour integrals in complexified cross-ratio space against a carefully chosen weight function. It was checked that the functionals coming from the numerical bootstrap seem to converge to the analytic expressions when the dimension of the numerical search space is increased towards infinity. 

An important obstacle to generalizations of this construction to more interesting situations has been the fact that in that work the knowledge of the exchanged spectrum in the extremal theory was used to constrain the functionals. In principle, we would like to do the opposite: derive the spectrum of an interacting CFT from the functional arising as the solution of an optimization problem. In this paper, we abstract the main features of the construction in \cite{Mazac:2016qev} without relying on a specific spectrum. We work with a broad class of functionals parametrized by a weight function which is holomorphic in an appropriate region of the complexified cross-ratio space. The action of the functional on the vectors corresponding to varying conformal families becomes a certain integral transform of the weight function. The search for an optimal bootstrap bound boils down to the search for the optimal weight function subject to constraints on its integral transform.

We will find the optimal weight function in two regimes. Firstly, we construct functionals that annihilate the spectrum of the generalized free fermion for general $\Delta_\phi > 0$. In this way, we demonstrate that the generalized free fermion maximizes the gap above identity for any $\Delta_\phi>0$. In this context, we also construct a related family of functionals which provide upper bounds on the OPE coefficient of a primary of dimension $2\Df+1$ and where the bounds are again saturated by the free fermion theory. The upper bound ceases to exist unless we impose a minimal gap on the spectrum that can be determined exactly for any $\Df$. This is a toy example of how our methods can predict nontrivial features in conformal bootstrap bounds. 

More importantly, we study the scaling limit of the 1D bootstrap equations where the dimensions of the external and exchanged operators become large with fixed ratios. This limit is relevant because it is equivalent to the large-radius limit of massive theories in $AdS$, as discussed in \cite{Paulos:2016fap}. Since $AdS$ in the limit of infinitely large radius becomes flat space, we expect that the corresponding limit of the 1D bootstrap equations should be related to the flat-space S-matrix bootstrap. We use the technique of analytic functionals to explain exactly how this happens.

Specifically, we consider the problem of finding the optimal upper bound on the OPE coefficient squared $c_{\phi\phi\mathcal{O}}^2$ of a primary operator $\mathcal{O}$ whose dimension satisfies $0<\Delta_{\mathcal{O}}<2\Delta_\phi$ provided all other primary operators in the $\phi\times\phi$ OPE have dimensions at least $2\Delta_\phi$. In the scaling limit, we take $\Df,\Delta_{\mathcal{O}}\rightarrow\infty$ with $m=\Delta_{\mathcal{O}}/\Df$ fixed. In the context of a massive field theory in $AdS_2$, this problem corresponds to bounding the coupling of a bound state $\mathcal{O}$ to its constituent $\phi$-particles, assuming this is the unique bound state that two $\phi$-particles form. The parameter $m$ has the meaning of the mass of $\mathcal{O}$ in units of the mass of $\phi$. This question has been solved analytically in flat space \cite{Creutz:1973rw} (see also \cite{Paulos:2016but}), the optimal answer being given by an exact S-matrix in the sine-Gordon theory, with $\phi$ and $\mathcal{O}$ the lightest and second-lightest breathers respectively. The corresponding 1D conformal bootstrap problem was studied numerically in \cite{Paulos:2016fap}, where substantial evidence was presented that the optimal solution to crossing in the $\Df\rightarrow\infty$ limit coincides with the boundary dual of the sine-Gordon theory in large $AdS_2$.

We use our construction of bootstrap functionals to prove this statement analytically on the 1D CFT side (without assuming any underlying $AdS$ description). The reason this is technically possible is that in the $\Df\rightarrow\infty$ limit, the action of our functionals localizes to a saddle point, revealing a direct relation between the conformal cross-ratio and the flat-space Mandelstam variable. Optimizing the weight function then leads to an extremal functional whose double zeros approach the operators corresponding to two-particle states of the sine-Gordon theory placed in large $AdS_2$. Indeed, the limiting optimal weight function is essentially the inverse of the sine-Gordon S-matrix. The extremal functional leads to the following upper bound on the OPE coefficient:
\be
c_{\phi\phi\mathcal{O}}^2 \leq
\sqrt{64\pi\Df}
\frac{m^{3/2}\sqrt{2-m}}{\left|m^2-2\right| \sqrt{2+m}}
\left[\frac{2^{2(m+2)}}{
\left(2-m\right)^{2-m}\left(2+m\right)^{2+m}}\right]^{-\Df}\,
\ee
valid asymptotically in the limit $\Df\to \infty$. The right-hand side agrees with the OPE coefficient squared in the holographic dual of the sine-Gordon theory in large $AdS_2$, further confirming our proposal. We also show how these results generalize to situations with multiple bound states.

The outline for the rest of this paper is as follows. In the next section, we describe the theories to which our results apply and review the general mechanism of how functionals lead to bounds on the CFT data. In section \ref{sec:funcs}, we motivate and define the class of functionals that we will be working with, spelling out the constraints that the weight function should satisfy. In section \ref{sec:free}, we explicitly solve these constraints in the case of the generalized free fermion, proving that this theory maximizes the gap above identity for any value of the external dimension $\Delta_\phi$. We also find the same theory saturates an upper bound on the OPE coefficient of an operator of dimension $2\Df+1$, and construct the associated functionals. Section \ref{sec:largeR} considers OPE maximization in the large-radius limit. We show that the resulting extremal functionals have spectra determined by S-matrices of two-dimensional theories with no particle production. We finish with conclusions and future prospects. Several appendices complement results in the main text.

\section{Review of 1D CFTs and bootstrap functionals}
\label{sec:review}
\subsection{1D CFTs}
We consider unitary theories invariant under $SO(1,2)$ realized as the 1D conformal group. Note that these are in general very different from one-dimensional Weyl-invariant theories (such as the ones studied in \cite{Mezei:2017kmw}), which are necessarily topological. The Lie algebra of $SO(1,2)$ is generated by operators $P$, $D$ and $K$ corresponding to an infinitesimal translation, dilatation and special conformal transformation respectively. The space of states of the radially-quantized theory carries a positive scalar product such that $D^\dagger=D$, $P^\dagger=K$. We assume this Hilbert space contains a vacuum state invariant under all three generators.\footnote{In particular, this assumption excludes standard conformal quantum mechanics studied in \cite{deAlfaro:1976vlx}.}

The standard state-operator map provides an isomorphism between this Hilbert space and the space of local operators that can be inserted on the line. Primary operators $\mathcal{O}(x)$ satisfy $[K,\mathcal{O}(0)] = 0$. The eigenvalue under the action of dilatations is the scaling dimension, denoted $\Delta$: $[D,\mathcal{O}(0)] = \Delta\,\mathcal{O}(0)$. An important constraint that follows from the positivity of the scalar product is that all dimensions should be non-negative. For our purposes, a 1D CFT is a set of correlation functions of local operators on the line satisfying reflection positivity and the global Ward identities following from $D,K,P$ and the existence of an invariant vacuum.

The set-up just described finds numerous interesting realizations in physics. A large class of such models consists of conformal boundaries and interfaces in 2D CFTs, or more generally conformal line defects in higher-D CFTs. Examples studied recently from the point of view of conformal field theory include the monodromy line defect in the 3D Ising model \cite{Billo:2013jda,Gaiotto:2013nva}, Wilson lines in four-dimensional $\mathcal{N}=4$ SYM \cite{Giombi:2017cqn,Beccaria:2017rbe,Giombi:2018qox} and Wilson lines in the ABJM theory \cite{Bianchi:2017ozk,Bianchi:2018scb}. A special case of this scenario is the trivial defect, where we simply restrict the operators of a higher-D CFT to lie on a line. The resulting 1D correlation functions satisfy all our assumptions. While we do not expect our bounds to be optimal in this case since they do not use the full conformal invariance, it is useful to keep in mind they are still valid.

Another set of examples particularly relevant for our analysis comes from a ``rigid'' (non-gravitational) version of $AdS_2$ holography, formulated in detail in \cite{Paulos:2016fap}. Consider any unitary UV-complete (1+1)D quantum field theory. We expect it is generally possible to place the theory in a non-dynamical $AdS_2$ background while preserving the $SO(1,2)$ group of isometries of $AdS_2$. In the process, we need to choose an $SO(1,2)$-invariant boundary condition on the bulk fields. The standard holographic dictionary \cite{Gubser:1998bc,Witten:1998qj} then gives a set of 1D correlation functions satisfying all global conformal Ward identities. The radially quantized Hilbert space of the boundary theory coincides with that of the bulk using equal-time slicing in global $AdS_2$. The existence of a finite $AdS$ radius $R$ gives us the possiblity to make various dimensionless couplings in the bulk Lagrangian into functions of $MR$, where $M$ is the mass-scale of the theory. In order to obtain a well-defined flat-space limit, we should restrict to situations where the couplings approach finite values as $R\rightarrow\infty$. We refer the reader to \cite{Paulos:2016fap} for more details on this version of the holographic correspondence.

Last but not least, there are genuinely 1D, intrinsically defined, non-trivial conformal-invariant theories. One example is the 1D long-range Ising model which can have a second order phase transition described by a non-trivial CFT \cite{Behan2017,Behan2017a,Paulos2016}. Another example which has recently received considerable attention is the SYK model \cite{1993PhRvL..70.3339S,Maldacena:2016hyu}. While conformal invariance is explicitly broken in the original model, it is possible to modify the UV kinetic term to produce a (non-local) model with exact conformal invariance such that it coincides with the SYK model at strong coupling \cite{Gross:2017vhb}.\footnote{Theories with $SO(1,2)$ invariance can also arise from non-relativistic theories with Schr\"{o}dinger symmetry \cite{Pal:2018idc}.}

\subsection{Crossing equations and linear functionals}
Having listed the kinds of theories to which our results will apply, let us describe the set-up for our analysis. We consider the OPE of a self-conjugate primary operator $\phi$ with itself. We make no a priori assumptions on the statistics of the field. The self-conjugate property implies the OPE contains the identity operator. The CFT data of the operators exchanged in the OPE is constrained by the crossing symmetry of the four-point function $\langle\phi(x_1)\phi(x_2)\phi(x_3)\phi(x_4) \rangle$. Without loss of generality, we can order the positions as $x_1<x_2<x_3<x_4$. Conformal invariance ensures that the correlator can be written as
\be
\langle\phi(x_1)\phi(x_2)\phi(x_3)\phi(x_4) \rangle = \frac{1}{|x_{12}x_{34}|^{2\Delta_\phi}}\mathcal{G}(z)\,,
\ee
where $x_{ij} = x_i - x_j$ and $\mathcal{G}(z)$ depends only on the cross-ratio
\be
z = \frac{x_{12}x_{34}}{x_{13}x_{24}}\,.
\ee
In the chosen ordering of positions we have $z\in (0,1)$. In this region, there are two convergent OPEs, namely the s-channel $x_2\rightarrow x_1$ and the t-channel $x_2\rightarrow x_3$, giving rise to the expansions
\ba
&\mathcal{G}(z) \stackrel{\textrm{s-channel}}{=}\sum\limits_{\mathcal{O}\in\phi\times\phi}\!\!c_{\phi\phi\mathcal{O}}^2 G_{\Delta_{\mathcal{O}}} (z)\\
&\mathcal{G}(z) \stackrel{\textrm{t-channel}}{=}\left(\frac{z}{1-z}\right)^{2\Delta_{\phi}}\!\!\sum\limits_{\mathcal{O}\in\phi\times\phi}\!\!c_{\phi\phi\mathcal{O}}^2 G_{\Delta_{\mathcal{O}}} (1-z)\,,
\ea
where the sums run over the primary operators exchanged in the $\phi\times\phi$ OPE, $c_{\phi\phi\mathcal{O}}$ are the OPE coefficients and $G_{\Delta}(z)$ the 1D conformal blocks
\be
G_{\Delta}(z) = z^{\Delta}{}_2F_1(\Delta,\Delta;2\Delta;z)\,.
\ee
In unitary theories, $c_{\phi\phi\mathcal{O}}$ are real so that $c_{\phi\phi\mathcal{O}}^2 \geq 0$. The equality of the two expansions above is a necessary and sufficient condition for the full crossing symmetry of the four-point function. It is convenient to rewrite the equality as
\be
\sum\limits_{\mathcal{O}\in\phi\times\phi}\!\!c_{\phi\phi\mathcal{O}}^2 F_{\Delta}^{\Delta_\phi}(z) = 0\,,
\label{eq:bootstrap1}
\ee
by defining
\be
F_{\Delta}^{\Df}(z) =z^{-2\Df}G_{\Delta}(z) - (1-z)^{-2\Df} G_{\Delta}(1-z)\,.
\label{eq:Fvector}
\ee
The variable $z$ is real and between 0 and 1 in the physical regime, but the region of validity of \eqref{eq:bootstrap1} is greater. Indeed, one can use the $\rho$ variable introduced in \cite{Hogervorst:2013sma} to show the s-channel expansion converges in the whole complex $z$-plane except for $z\in[1,\infty)$.\footnote{We should keep in mind that the 1D s-channel conformal blocks have a branch cut starting at $z=0$. The more precise statement then is that the s-channel OPE expansion converges on a multi-sheet cover of $\mathbb{C}\backslash[1,\infty)$.} Correspondingly, the t-channel expansion converges in the complex $z$-plane except for $z\in(-\infty,0]$. Therefore, equation \eqref{eq:bootstrap1} is valid for $z\in\mathcal{R}\equiv\mathbb{C}\backslash((-\infty,0]\cup[1,\infty))$. In particular, this means we are not allowed to analytically continue the equation through either of the branch cuts, and have to stay on the first sheet.
\begin{figure}[ht!]%
\begin{center}
\includegraphics[width=8cm]{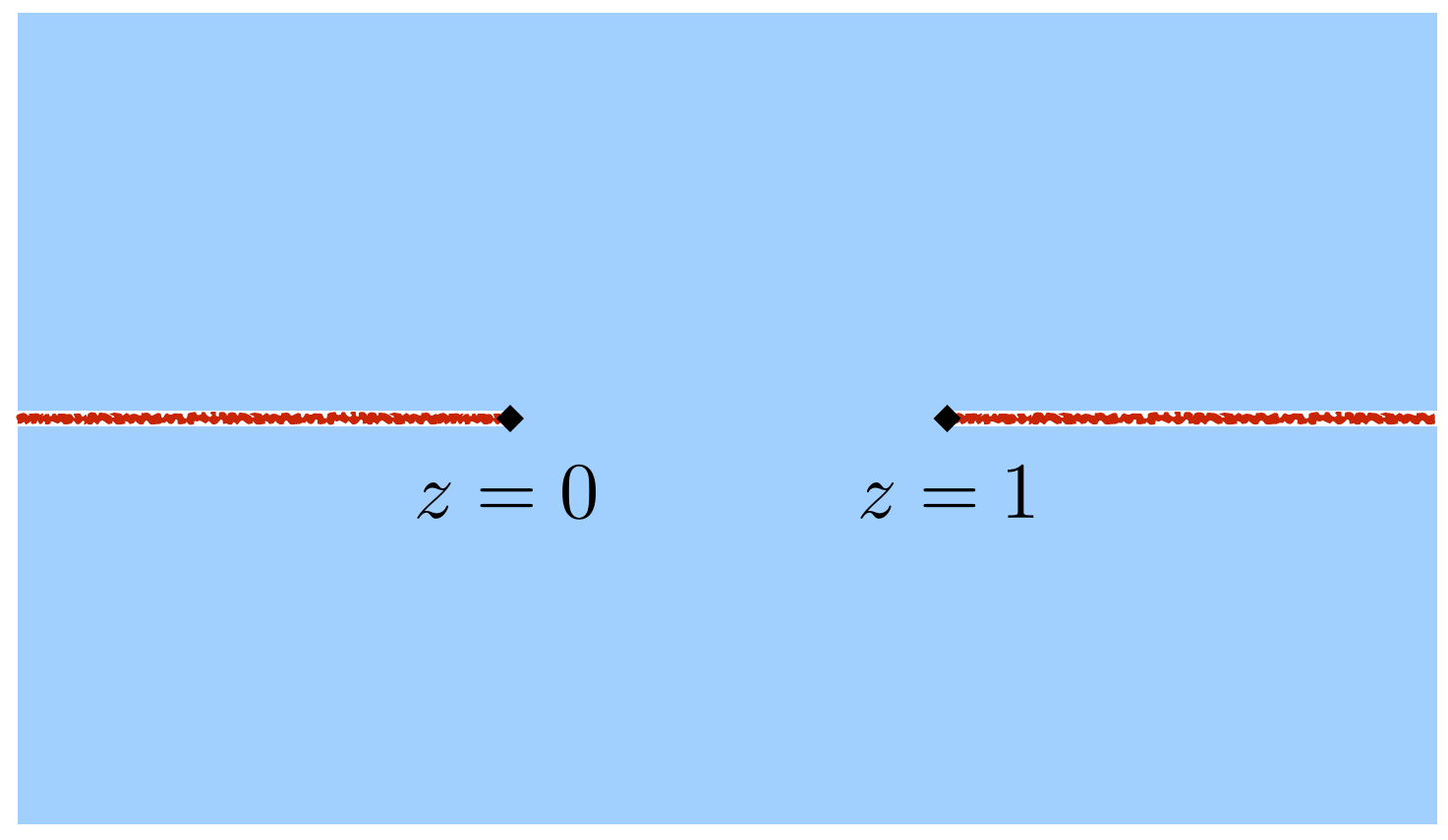}%
\caption{The physical region where the 1D crossing equation holds is $z\in(0,1)$. Allowing complex $z$ shows the full domain of validity of the equation is the entire blue region
$\mathcal{R}\equiv\mathbb{C}\backslash((-\infty,0]\cup[1,\infty))$. The equation stops holding after an analytic continuation from this region through one of the branch cuts.}
\label{fig:regionR}%
\end{center}
\end{figure}

The bootstrap equation \eqref{eq:bootstrap1} naturally lives in a vector space of functions of variable $z$, antisymmetric under $z\mapsto 1-z$ and analytic in $\mathcal{R}$. It is thus useful to parametrize the bootstrap constraints by elements of the dual space, usually called functionals. Given one such functional, denoted $\omega$, we can apply it to \eqref{eq:bootstrap1} and get a sum rule constraining the CFT data
\be
\sum\limits_{\mathcal{O}\in\phi\times\phi}\!\!c_{\phi\phi\mathcal{O}}^2\,\omega(\Delta_{\mathcal{O}}) = 0\,,
\label{eq:bootstrap2}
\ee
where the function $\omega(\Delta)$ is defined as the action of $\omega$ on the vectors $F_{\Delta}^{\Df}(z)$
\be
\omega(\Delta) \equiv \omega\left[F_{\Delta}^{\Df}\right].
\ee
The dependence on $\Df$ is supressed for simplicity in this notation. The simplest examples of functionals are evaluations of a function or its derivatives at a point in $\mathcal{R}$. The numerical bootstrap typically uses odd-order derivatives evaluated at $z=1/2$ as a basis for functionals. Any sum rule of the form \eqref{eq:bootstrap2} which is valid on any physical solution to crossing should arise as an action of some functional supported in $\mathcal{R}$.

It is important to note that not every functional which is linear when acting on finite linear combinations of functions can be used to infer the sum rule \eqref{eq:bootstrap2}. The reason is that in deriving \eqref{eq:bootstrap2} from \eqref{eq:bootstrap1}, we have to exchange the action of $\omega$ with an \emph{infinite} sum over operators appearing in the OPE. Reference \cite{Rychkov:2017tpc} formulated necessary and sufficient conditions that a functional $\omega$ should satisfy in order for equation \eqref{eq:bootstrap2} to be valid for any physical solution to crossing \eqref{eq:bootstrap1}. This property, called the \emph{swapping condition}, is automatic if the functional only depends on values of test functions at a finite distance away from the boundary of $\mathcal{R}$, thanks to the exponentially fast convergence of the OPE~\cite{Pappadopulo:2012jk}. However, it turns out that the analytic functionals constructed in \cite{Mazac:2016qev} and later in this work necessarily depend on values of test functions arbitrarily close to the boundary of $\mathcal{R}$. Therefore, the conditions spelled out in \cite{Rychkov:2017tpc} will play an important role here. For completeness, we review the implications of the swapping condition in appendix \ref{app:falloff}.

\subsection{Optimization and bounds}
\label{sec:optbounds}
A bootstrap question that is particularly natural from the functional perspective is the problem of {\em gap maximization}: we would like to know what is the maximal allowed value for the dimension of the first, non-identity operator in the OPE $\phi\times \phi$, among unitary solutions to \eqref{eq:bootstrap1}.  The key idea of \cite{Rattazzi:2008pe} is that we get an upper bound on this gap value by constructing a linear functional satisfying
\be
\omega(0)>0\,, \qquad \forall \Delta\geq \Delta^*:\, \omega(\Delta)\geq 0.
\ee
It follows from applying this functional to the crossing equation \eqref{eq:bootstrap1} that the gap is at most $\Delta^*$. The maximal value of the gap allowed by crossing symmetry and unitarity is the infimum of values of $\Delta^*$ for which such a functional exists. Accordingly, the limiting functional is called the optimal or {\em extremal} functional~\cite{ElShowk:2012hu}. It is generically unique (up to rescaling) and has the following properties
%
%
\be
\omega(0)=0\,, \qquad \omega(\Delta_n)=0\,,\qquad
\omega'(\Delta_n)\propto\delta_{n,0}\,,
\ee
where $\{\Delta_n:n\in\mathbb{Z}_{\geq 0}\}$ is a discrete, increasing set of scaling dimensions such that $\Delta_0=\Delta^*$. This set is expected to gives rise to a solution to crossing maximizing the gap,\footnote{Strictly speaking, to our knowledge this has only been proved fully rigorously  (see e.g. \cite{Reemtsen1998}) for truncations of the problem we are considering here, where the continous set of crossing constraints are reduced to a finite dimensional set. However, this does not have any effect on the validity of our arguments in this paper.} i.e. there exist $c_n^2\geq 0$ such that
\bea
F_0^{\Df}(z)+\sum_{n=0}^{\infty} c_n^2 F_{\Delta_n}^{\Df}(z)=0.
\eea
In Section \ref{sec:free}, we will construct the extremal functionals for the gap maximization problem in 1D analytically.

Another interesting bootstrap question readily addressed using functionals is OPE maximization~\cite{Rattazzi2011}. In this case we fix $\Delta_b$ of an exchanged operator $\mathcal{O}_b$ whose OPE coefficient we want to bound, and assume the rest of the spectrum besides identity lies in some set $S$. The crossing sum rule is then
\be
F_0^{\Df}(z)+c_{b}^2\, F_{\Db}^{\Df}(z)+\sum_{\Delta \in S} c_{\Delta}^2 F_{\Delta}^{\Df}(z)=0\,.
\ee
for some real valued\footnote{The summation notation is schematic. It should be replaced by an integral, with $c_\Delta$ a real-valued distribution with support in $S$ satisfying 
\bea
\int_{\Delta-\epsilon}^{\Delta+\epsilon} \!\!\ud \Delta'\, c_{\Delta'}^2 \geq 0\qquad \mbox{for all} \quad \epsilon>0, \Delta\in S.
\eea
That is, our results will apply to general solutions to crossing including those involving a continuum of operators in the OPE.
}
$c_\Delta$. 
We can obtain an upper bound on $c_{b}^2$ by constructing a functional satisfying
\bea
\omega(\Delta_b)>0, \qquad \forall\Delta \in S:\, \omega(\Delta)\geq 0.
\eea
Indeed, applying such a functional to the sum rule we obtain
\bea
c_b^2 \leq -\frac{\omega(0)}{\omega(\Delta_b)}.
\eea
In the extremal case, this inequality is saturated and the extremal functional satisfies
\bea
c_{b,\textrm{max}}^2=-\frac{\omega(0)}{\omega(\Delta_b)}, \qquad \forall n\geq0:\,\omega(\Delta_n)=0,\qquad
\forall n\geq1:\,\partial_{\Delta}\omega(\Delta_n)=0\,,
\eea
where again $\Delta_n$ is an increasing sequence of scaling dimensions forming the spectrum of the extremal solution, i.e. there exist $c_n^2 >0$ so that
\bea
F_0^{\Df}(z)+c_{b,\textrm{max}}^2\, F_{\Db}^{\Df}(z)+\sum_{n=0}^{\infty} c_n^2 F_{\Delta_n}^{\Df}(z)=0\,.
\eea
In general, both the bound and the solution will depend on the original choice of set $S$.

The main lesson is that in both the gap and OPE maximization problems, the knowledge of the extremal functional immediately gives us the spectrum of the extremal solution to crossing. It is clear then that it is crucial to understand the exact form taken by extremal functionals. In numerical bootstrap studies, one truncates the space of functionals to that generated by finitely many derivatives with respect to cross-ratios evaluated at the crossing-symmetric point. As the number of derivatives is increased, the bounds monotonically improve and converge to the optimal ones. The action of the optimal numerical functionals on $F_{\Delta}^{\Df}$, i.e. $\omega(\Delta)$ also converges as the number of derivatives increases, giving us better and better approximations to the exact spectrum. However, the coefficients of the $z$-derivatives generically do not converge, indicating that the exact extremal functionals themselves may not be expressible in terms of $z$-derivatives.

In the following section, we discuss a very general class of functionals taking the form of a contour integral in complexified cross-ratio space against a holomorphic weight function. In this language, we will be able to fix various extremal functionals analytically and use them to derive non-trivial spectra of extremal solutions to crossing.

\section{General Functionals}
\label{sec:funcs}
\subsection{Functionals as contour integrals}
\label{ssec:funcsGen}
As discussed in the previous sections, bootstrap functionals act on functions analytic in $\mathcal{R} = \mathbb{C}\backslash((-\infty,0]\cup[1,\infty)$ and satisfying $\mathcal{F}(z) = - \mathcal{F}(1-z)$. The functionals must be linear when acting on finite linear combinations of test functions. The functionals must also take finite values on all bootstrap vectors $F_{\Delta}^{\Df}(z)$ with $\Delta\geq0$. Finally, the action of the functionals must be interchangeable with the infinite sum over primary operators in any unitary solution to crossing. Following \cite{Rychkov:2017tpc}, we will refer to the last two conditions as \emph{finiteness} and \emph{swapping}.

Let us begin by stating the form of functionals that we will consider. Later in this section, we will explain why this is a useful way to write a very general functional for the 1D crossing equation, although this is probably not apparent at first sight.
\be
\boxed{
\omega[\mathcal{F}] =
\frac{1}{2}\!\!\!\int\limits_{\frac{1}{2}}^{\frac{1}{2}+i\infty}\!\!\!\!dz\,f(z)\mathcal{F}(z) + 
\!\!\int\limits_{\frac{1}{2}}^{1}\!\!dz\,g(z)\mathcal{F}(z)}
\,.
\label{eq:ffg}
\ee
\begin{figure}[ht!]%
\begin{center}
\includegraphics[width=9cm]{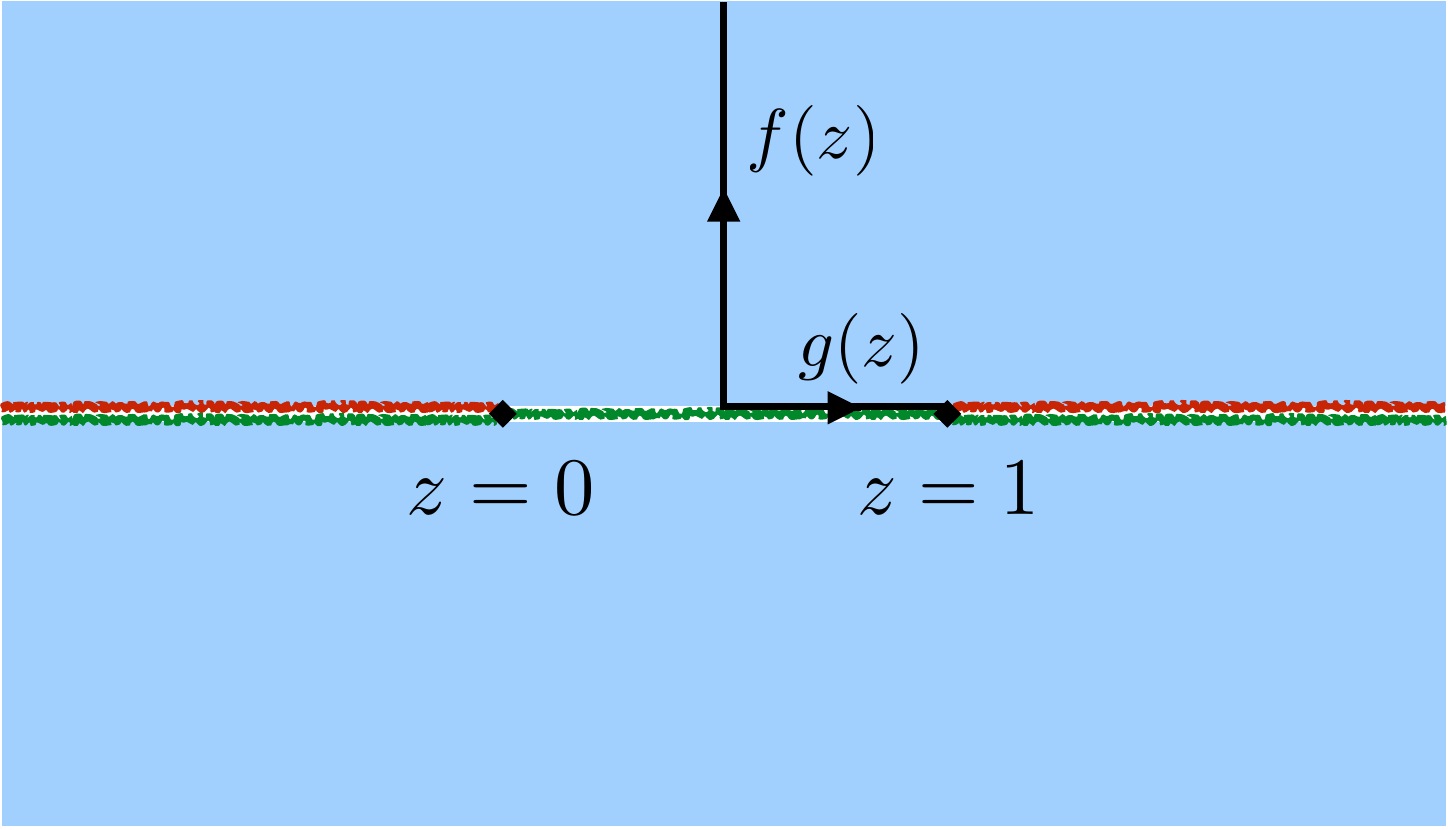}%
\caption{The representation of a general functional in terms of a pair of weight functions $f(z)$ and $g(z)$. The branch cuts of $f(z)$ are shown in green. The branch cuts of $g(z)$ coincide with those of the bootstrap vectors $F_{\Delta}^{\Df}(z)$ and are shown in red. The integration contours approach the boundary of $\mathcal{R}$ at $z=\infty$ and $z=1$, and consequently $f(z)$ and $g(z)$ respectively must have appropriate fall-off there.}
\label{fig:fgContour}%
\end{center}
\end{figure}
The functional takes the form of a pair of contour integrals inside $\mathcal{R}$, specified by weight functions $f(z)$ and $g(z)$. The factor of $1/2$ is inserted in the first integral for future convenience. To ensure the reality of the functional action on the bootstrap vectors, $f(z)$ and $g(z)$ should be real on the respective integration contours. We will assume $f(z)$ is complex-analytic in the upper-half plane, and that its only singularities on the real axis can be located at $z=0,1$. We will take $g(z)$ to be complex-analytic in $\mathcal{R}$. Furthermore, as we review in appendix \ref{app:falloff}, the functional only satisfies the finiteness and swapping conditions if the weight functions have the following fall-off properties in the region where the integration contours approach the boundary of $\mathcal{R}$
\bea
\exists\, \epsilon>0:\qquad f(z) \stackrel{|z|\to \infty}{\sim} o(z^{-1-\epsilon}), \qquad g(z)\stackrel{z\to 1}{\sim}o\left((1-z)^{2\Df-1+\epsilon}\right)\,, \label{eq:fgfalloff}
\eea
where the first condition applies for $\mathrm{Im}(z)>0$. It will be convenient to define $f(z)$ in the lower half plane in terms of its values in the upper half plane by setting $f(z)=f(1-z)$. Note that this implies such $f(z)$ may have a discontinuity on the real axis. When referring to $f(z)$ for $z\in\mathbb{R}$, we will always mean the value of $\lim_{\epsilon\rightarrow 0^+}f(z+i\epsilon)$. As we will see shortly, $f(z)$ and $g(z)$ both arise from the same complex-analytic function. The existence of this underlying analytic function guarantees the validity of our final constraint, which we will call the {\em gluing condition}
\be
\mathrm{Re}[f(z)] + g(z) + g(1-z) = 0\quad\textrm{for }z\in(0,1)\,.
\label{eq:glue}
\ee
In the rest of this section, we will explain in detail why this particular ansatz is a well-motivated starting point. The reader who is only interested in applications may however now safely skip ahead to later sections, where we will construct pairs of kernels $f,g$ which correspond to various interesting extremal functionals.

The proposal \eqref{eq:ffg} naturally arises from a description of functionals introduced in \cite{Mazac:2016qev}, which takes the following form:
\be
\omega[\mathcal{F}] = \frac{1}{2\pi i}
\!\int\limits_1^{\infty}\!\!dz\,h(z)\,\mathrm{Disc}[\mathcal{F}(z)]\,,
\label{eq:fh}
\ee
where
\be
\mathrm{Disc}[\mathcal{F}(z)] =\lim_{\epsilon\rightarrow 0^+}\left[\mathcal{F}(z+i\epsilon) - \mathcal{F}(z-i\epsilon)\right]\,.
\ee
Let us first demonstrate that this prescription efficiently encodes a very broad class of functionals. We will start with the simplest functional, namely the evaluation at a fixed point $w\in\mathcal{R}$:
\be
E_w[\mathcal{F}] = \mathcal{F}(w)\,.
\ee
It is clear that $E_w$ satisfies the finiteness and swapping condition, the latter by virtue of the (exponentially fast) convergence of both the s- and t-channel OPEs at $w$. In order to rewrite $E_w$ as \eqref{eq:fh}, we first write it as a contour integral
\be
E_w[\mathcal{F}] = \frac{1}{2\pi i}\oint\limits_C\!dz\,\frac{\mathcal{F}(z)}{z-w}\,,
\ee
where $C$ is a small circle around $w$. We can now pull the contour away from $w$ and wrap it around the branch cuts. When acting on the bootstrap vectors, there is no contribution from the arcs at infinity since
\be
F_{\Delta}^{\Df}(z) = O(z^{-2\Df}\log z)\,\textrm{ as }z\rightarrow\infty\,.
\ee
and $\Df>0$ by unitarity. Using the antisymmetry of $\mathcal{F}$ under crossing, we can combine the contribution from the two branch cuts and write the action of $E_w$ on all functions of interest as \eqref{eq:fh} with
\be
h(z) = \frac{2w-1}{(z-w)(z-1+w)}\,.
\label{eq:hEw}
\ee
We should now clarify a subtlety in the prescription \eqref{eq:fh}. When acting on the bootstrap vectors $F_{\Delta}^{\Df}$ with sufficiently small $\Delta$, there can be a divergence in the integral in \eqref{eq:fh} coming from $z\rightarrow1$, seemingly violating the finiteness condition. With $h(z)$ given by \eqref{eq:hEw} this happens for $\Delta<2\Df - 1$. However, we can pull the contour of integration away from the branch cut around $z=1$ to manifest finiteness. For general $h(z)$ we should demand that a similar contour deformation should be possible in order to make the result manifestly finite. The precise condition will be explained below.

Having seen how to express the evaluation functionals as \eqref{eq:fh}, let us write the most general functional as an integral of $E_w$ against a distribution $\rho(w,\bar{w})$
\be
\omega =\!\! \int\limits_{\mathcal{R}}\!\!d^2w\,\rho(w,\bar{w}) E_w\,,
\label{eq:fRho}
\ee
However, not every $\rho(w,\bar{w})$ leads to a functional satisfying finiteness and swapping. The simplest way to ensure that \eqref{eq:fRho} does satisfy finiteness and swapping is by restricting $\omega$ to only involve derivatives of a bounded order, integrated over a compact region in $\mathcal{R}$. We will refer to these as \emph{functionals of the first kind}. Functionals of the first kind are always consistent thanks again to the exponentially fast convergence of both OPEs in any compact subregion of $\mathcal{R}$. The functionals used in the numerical bootstrap are linear combinations of derivatives of bounded order evaluated at $z=1/2$, and are therefore of the first kind.

Every functional of the first kind can be expressed as \eqref{eq:fh} for a suitable $h(z)$. To see that, we use \eqref{eq:hEw} to write the action of $\omega$ on a test function $\mathcal{F}(z)$ as
\be
\omega[\mathcal{F}] =\frac{1}{2\pi i}\!\int\limits_{\mathcal{R}}\!\!d^2w\,\rho(w,\bar{w})
\!\int\limits_1^{\infty}\!\!dz\frac{2w-1}{(z-w)(z-1+w)}\mathrm{Disc}[\mathcal{F}(z)]\,.
\ee
For functionals of the first kind, the two integrations can be swapped. This is because the support of $\rho(w,\bar{w})$ stays away from the singularities of the kernel \eqref{eq:hEw}, and is bounded from $w=\infty$. Therefore, $\omega$ is of the form \eqref{eq:fh} with
\be
h(z) = \!\int\limits_{\mathcal{R}}\!\!d^2w\,\frac{2w-1}{(z-w)(z-1+w)}\rho(w,\bar{w})\,.
\label{eq:hInt}
\ee
Note that with $h(z)$ given by this formula, the form $h(z) dz$ is holomorphic in an open neighbourhood $U$ of $\partial\mathcal{R}$ including $z=\infty$. Furthermore, $h(z) = h(1-z)$ in $U$. If we wish, we can use this symmetry to double the contour in \eqref{eq:fh} on the other branch cut and then deform it to the interior of $\mathcal{R}$, obtaining a functional that manifestly satisfies finiteness and swapping. In summary, we see that there is a one-to-one correspondence between functionals of the first kind and functions $h(z)$ satisfying the properties stated after \eqref{eq:hInt}.\footnote{Strictly speaking, we did not demonstrate that the map from functions $h(z)$ with the required properties to functionals is injective, but this seems to be the case.} For example, when $\omega$ is a finite linear combination of derivatives at $z=1/2$
\be
\omega[\mathcal{F}] = \sum\limits_{j=1}^{N}\frac{\alpha_j}{(2j-1)!}\partial^{2j-1}\mathcal{F}(w)|_{w=\frac{1}{2}}\,,
\ee
the corresponding weight function becomes
\be
h(z) = 2\sum\limits_{j=1}^{N}\frac{\alpha_j}{\left(z-\frac{1}{2}\right)^{2j}}\,.
\ee

It turns out that there are interesting consistent functionals which are not of the first kind. Such functionals either involve taking derivatives of arbitrarily high order, or such that the support of $\rho(w,\bar{w})$ reaches all the way to the boundary $\partial\mathcal{R}$.\footnote{In fact, there is no clear distinction between these two cases since the test functions are holomorphic, and so their behaviour on $\partial\mathcal{R}$ is encoded in their derivatives at an interior point.} We will refer to all consistent functionals which are not of the first kind as \emph{functionals of the second kind}. Functionals of the second kind are of fundamental importance for the conformal bootstrap. The extremal functionals for the gap and OPE maximization problems constructed in \cite{Mazac:2016qev} and in the following sections of this article are all of the second kind. There is also some evidence that the extremal functionals in higher dimensions probe the light-cone limit and are therefore of the second kind too \cite{Simmons-Duffin:2016wlq}.

\begin{figure}[ht!]%
\begin{center}
\includegraphics[width=16cm]{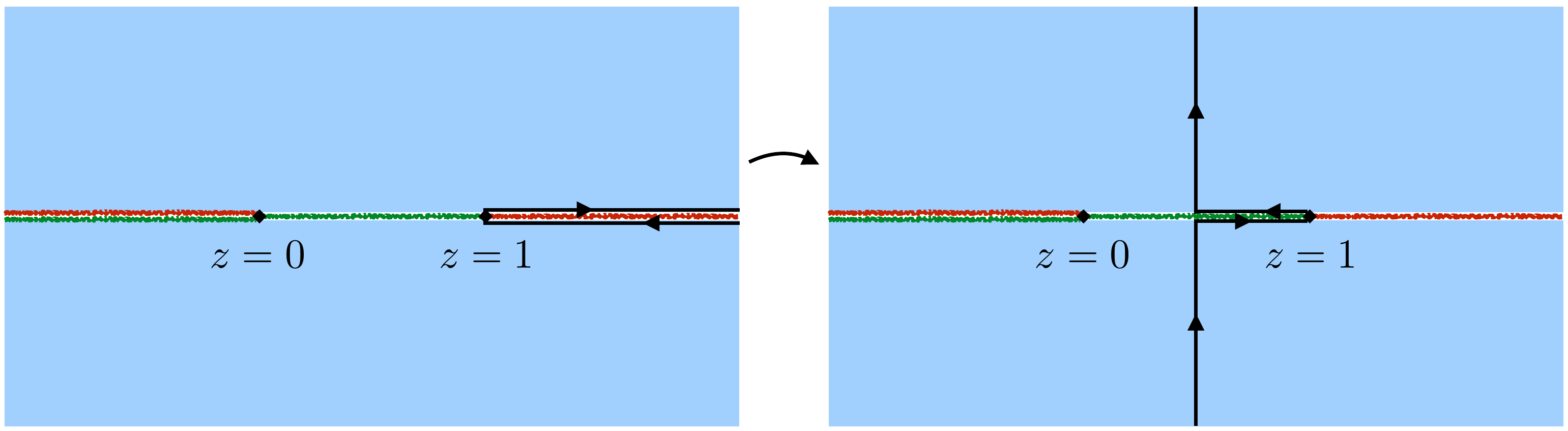}%
\caption{The contour deformation we can use to go from the representation of a general functional \eqref{eq:fh} in terms of $h(z)$ to the more convenient representation \eqref{eq:ffg} in terms of $f(z)$ and $g(z)$. The branch cuts of $F_{\Delta}^{\Df}(z)$ are shown in red and the branch cuts of $h(z)$ are shown in green. Whenever we have two integration contours running in opposite directions above and below a branch cut, we are really integrating the discontinuity across the branch cut.}
\label{fig:contourDef1}%
\end{center}
\end{figure}

The key point is that the prescription \eqref{eq:fh} also includes numerous functionals of the second kind. In light of the preceding discussion, they correspond to $h(z)$ not analytic in any open neighbourhood of $\partial\mathcal{R}$. While it would be interesting to see if every consistent functional of the second kind can be represented by \eqref{eq:fh} for suitable $h(z)$, this question is beyond the scope of this work. In a general functional, the analytic structure of $h(z)$ can be very complicated. However, we will see that in order to describe the extremal functional of interest, it will be enough to focus on $h(z)$ with nice analytic properties. We will first postulate these properties and later see why they make sense. These properties are:
\ba
1.\quad &h(z)\textrm{ is analytic away from possible poles or branch points at }z=0,1\textrm{ and }\infty.\\
2.\quad &h(z)\textrm{ is bounded by }A_1 |z|^{-1-\epsilon_1}\textrm{ for some }A_1,\epsilon_1>0\textrm{ as }z\rightarrow\infty.
\\
3.\quad &\textrm{The discontinuity of }h(z)\textrm{ around }z=1\textrm{ is bounded by }A_2 |z-1|^{2\Df - 1 + \epsilon_2}\\
&\textrm{ for some }A_2,\epsilon_2>0\textrm{ as }z\rightarrow1.
\label{eq:hProperties}
\ea
Property 1 in particular means that $h(z)$ can be analytically continued through possible branch cuts stretching between $z=0,1$ and $\infty$ without encountering any other non-analyticities.\footnote{It is easy to see that every $h(z)$ with these properties describes a functional of the second kind. Property 2 implies $h(z)$ cannot be an entire function. Property 1 then implies either $z=1$ or $z=\infty$ is a singular point of $h(z)$, and hence the functional cannot be of the first kind.} Our first order of business is to show that a functional with the above properties satisfies finiteness and swapping. Let us first deform the contour in \eqref{eq:fh} into the interior of $\mathcal{R}$ as shown in figure \ref{fig:contourDef1}.

In this way we arrive back at our original representation \eqref{eq:ffg}, which we repeat here for convenience:
\be
\omega[\mathcal{F}] =\frac{1}{2}
\!\!\!\int\limits_{\frac{1}{2}}^{\frac{1}{2}+i\infty}\!\!\!\!dz\,f(z)\mathcal{F}(z) + 
\!\!\int\limits_{\frac{1}{2}}^{1}\!\!dz\,g(z)\mathcal{F}(z)
\,.
\ee
The weight functions $f(z)$ and $g(z)$ are now determined in terms of $h(z)$
\ba
f(z) &= \frac{h(z) - h(1-z)}{\pi i}\quad\textrm{for }\mathrm{Im}(z)>0\\
g(z) &= -\frac{\mathrm{Disc}[h(z)]}{2\pi i}\quad\quad\,\,\;\textrm{for }z\in(0,1)\,.
\label{eq:fgDef}
\ea
Without loss of generality, we can assume $h(z)$ is real for $z\in(1,\infty)$, so that the functional takes real values on the bootstrap vectors.\footnote{A complex-valued $h(z)$ can be written as $h_1(z) + i h_2(z)$, where $h_{1}(z) = [h(z) + \overline{h(\bar{z})}]/2$, $h_{2}(z) = [h(z) - \overline{h(\bar{z})}]/(2i)$. $h_{1,2}(z)$ are real for $z\in(1,\infty)$ and separately satisfy properties \eqref{eq:hProperties}.} It follows that $h(\bar{z}) = \overline{h(z)}$. This in turn implies
\be
f(z) = \overline{f(1-\bar{z})}\;\textrm{ for }\mathrm{Im}(z)>0\quad\textrm{and}\quad
g(z)\in\mathbb{R}\;\textrm{ for }z\in(0,1)\,.
\ee
It will be convenient for future calculations to extend the definition of $f(z)$ to $\mathrm{Im}(z)<0$ via
\be
f(z) \equiv f(1-z)\quad\textrm{for }\mathrm{Im}(z)<0\,.
\label{eq:fExtend}
\ee
At the same time, we will define $g(z)$ for $z\in\mathcal{R}$ by the analytic continuation from $z\in(0,1)$. The reality conditions on $f(z)$ and $g(z)$ then read
\be
f(\bar{z}) = \overline{f(z)}\quad\textrm{and}\quad
g(\bar{z}) = \overline{g(z)}\,.
\label{eq:fgReality}
\ee
In particular, $\mathrm{Re}[f(z)]$ is continuous across the real axis. As a result of \eqref{eq:fgDef}, $f(z)$ and $g(z)$ are constrained by the gluing condition mentioned previously,
\be
\mathrm{Re}[f(z)] + g(z) + g(1-z) = 0\quad\textrm{for }z\in(0,1)\,.
\ee

It is manifest from the form of the functional action \eqref{eq:ffg} that the only potential sources of violation of finiteness and swapping are $z\rightarrow\infty$ in the first integral and $z\rightarrow 1$ in the second integral. $|f(z)|$ is bounded by $A_1 |z|^{-1-\epsilon_1}$ for some $A_1,\epsilon_1>0$ as $z\rightarrow \infty$ by property 2, which is sufficient (and necessary) for finiteness and swapping of the first integral. Similarly, $|g(z)|$ is bounded by $A_2 |z-1|^{2\Df-1+\epsilon_2}$ for some $A_2,\epsilon_2>0$ as $z\rightarrow1$ by property 3, which is sufficient (and necessary) for finiteness and swapping of the second integral. Both these statements were demonstrated by \cite{Rychkov:2017tpc} and are reviewed in appendix \ref{app:falloff}. We conclude that both \eqref{eq:ffg} and equivalently \eqref{eq:fh} with the above mentioned conditions define consistent bootstrap functionals.

To close this section, let us mention that given $f(z)$ and $g(z)$ satisfying the mentioned constraints, we may easily recover the associated $h(z)$ from \eqref{eq:hInt}, finding
\be
h(z) = \frac{1}{2}\!\!\!\int\limits_{\frac{1}{2}}^{\frac{1}{2}+i\infty}\!\!\!\!dw\,\frac{2w-1}{(z-w)(z-1+w)}f(w)+ 
\!\!\int\limits_{\frac{1}{2}}^{1}\!\!dw\,\frac{2w-1}{(z-w)(z-1+w)}g(w)\,.
\ee
More precisely, this equation defines $h(z)$ in the region $\mathrm{Re}(z)>1/2$, $z\notin(1/2,1)$. Analytically continuing this function to the rest of the upper and lower half plane, we find $h(z)$ satisfying properties \eqref{eq:hProperties}. Notice that the gluing condition is equivalent to the statement that $h(z)$ has no discontinuity around $z=1/2$, i.e. property 1 holds.

\subsection{Simplifying the functional action}\label{ssec:funcSimp}
Having seen that the class of functionals defined by \eqref{eq:ffg}, or equivalently \eqref{eq:fh}, are consistent bootstrap functionals under suitable conditions on $f(z),g(z)$ or $h(z)$, let us explain why it is also naturally adapted to tackle bootstrap problems. To do this, we need to understand the action of \eqref{eq:fh} on the bootstrap vectors $F_{\Delta}^{\Df}(z)$. We start with the representation \eqref{eq:fh} and change variables in the term involving the direct-channel block (recall that $\omega(\Delta)\equiv\omega(F_{\Delta}^{\Df})$)
\ba
\omega(\Delta)&=  \frac{1}{2\pi i}
\!\int\limits_1^{\infty}\!\!dz\,h(z)\,\mathrm{Disc}\!\left[\frac{G_{\Delta}(z)}{z^{2\Df}}-\frac{G_{\Delta}(1-z)}{(1-z)^{2\Df}}\right]=\\
& =-\frac{1}{2\pi i}
\!\int\limits_{-\infty}^{0}\!\!dz\,h(1-z)\,\mathrm{Disc}\!\left[\frac{G_{\Delta}(1-z)}{(1-z)^{2\Df}}\right]-
\frac{1}{2\pi i}
\!\int\limits_1^{\infty}\!\!dz\,h(z)\,\mathrm{Disc}\!\left[\frac{G_{\Delta}(1-z)}{(1-z)^{2\Df}}\right]\,.
\ea
Note that $\mathrm{Disc}$ generates a minus sign under this change of variables. Our goal will be to bring all integrals to the region $z\in(1,\infty)$. The second term is already in this form. In the first term, we can first rotate the contours to the region $z\in(0,\infty)$ to find
\ba
&-\frac{1}{2\pi i}
\!\int\limits_{-\infty}^{0}\!\!dz\,h(1-z)\,\mathrm{Disc}\!\left[\frac{G_{\Delta}(1-z)}{(1-z)^{2\Df}}\right] = 
\!\!\int\limits_{0}^{1}\!\!dz\,g(1-z)\frac{G_{\Delta}(1-z)}{(1-z)^{2\Df}} 
+\\
&+\frac{1}{2\pi i}
\!\int\limits_{1}^{\infty}\!\!dz\left[e^{-i\pi(\Delta-2\Df)}h(1-z-i\epsilon)-e^{i\pi(\Delta-2\Df)}h(1-z+i\epsilon)\right]\frac{\widehat{G}_{\Delta}(1-z)}{(z-1)^{2\Df}}\,,
\label{eq:fMan}
\ea
where $g(z)$ is defined in \eqref{eq:fgDef}, and
\be
\widehat{G}_{\Delta}(z) = (-z)^{\Delta}{}_2F_1(\Delta,\Delta;2\Delta;z)\,.
\ee
In order to transform the first integral on the rhs of \eqref{eq:fMan} to the region $z\in(1,\infty)$, we perform the change of variables $z\mapsto1/z$ and use the following transformation of the conformal blocks
\be
G_{\Delta}(z) = \widehat{G}_{\Delta}\left(\frac{z}{z-1}\right)\,.
\ee
We find
\be
\int\limits_{0}^{1}\!\!dz\,g(1-z)\frac{G_{\Delta}(1-z)}{(1-z)^{2\Df}} = 
\int\limits_{1}^{\infty}\!\!dz\,z^{2\Df - 2}\,g\!\left(\frac{z-1}{z}\right)\frac{\widehat{G}_{\Delta}(1-z)}{(z-1)^{2\Df}} \,.
\ee
Combining the steps, we arrive at the result
\be
\omega(\Delta) = \int\limits_{1}^{\infty}\!\!dz\!\!\left[z^{2\Df - 2}\,g\!\left(\frac{z-1}{z}\right) -\frac{e^{-i\pi(\Delta-2\Df)}f(z) + e^{i\pi(\Delta-2\Df)}\overline{f(z)}}{2}\right]\frac{\widehat{G}_{\Delta}(1-z)}{(z-1)^{2\Df}} \,,
\label{eq:fAction}
\ee
with $f(z)$ and $g(z)$ defined in \eqref{eq:fgDef}. The values of $f(z)$ on the branch cut $(1,\infty)$ are obtained as the limit from $\mathrm{Im}(z)>0$ and the reality property \eqref{eq:fgReality} was used to get the last term in the square bracket. We can see the rhs of \eqref{eq:fAction} is manifesly real.

It is now natural to define integral transforms of $f(z)$ and $g(z)$ taking them to functions of $\Delta$ as follows\footnote{A similar integral transform has been introduced in \cite{Hogervorst2017a}. It would be interesting to explore possible connections between that work and our approach.}
\ba
\mathfrak{f}(\Delta)&\equiv
\!\!\int\limits_{1}^{\infty}\!\!dz\,f(z) \frac{\widehat{G}_{\Delta}(1-z)}{(z-1)^{2\Df}}\,,\\
\mathfrak{g}(\Delta)&\equiv
\!\!\int\limits_{1}^{\infty}\!\!dz\,z^{2\Df-2}g\!\left(\frac{z-1}{z}\right) \frac{\widehat{G}_{\Delta}(1-z)}{(z-1)^{2\Df}} = \!\!\int\limits_{0}^{1}\!\!dz\,g(z) \frac{G_{\Delta}(z)}{z^{2\Df}}\,.
\label{eq:intT}
\ea
The action of the functional \eqref{eq:fAction} can now be written compactly as
\be
\boxed{
\omega(\Delta) = \mathfrak{g}(\Delta) - \mathrm{Re}\!\left[e^{-i\pi(\Delta-2\Df)}\mathfrak{f}(\Delta)\right]}\,.
\label{eq:fActionT}
\ee
It is important to keep in mind that although $\omega(\Delta)$ is finite for any $\Delta\geq0$, the integral in \eqref{eq:fAction} generally converges only for $\Delta$ larger than some $\Delta_0$. For $0\leq\Delta\leq\Delta_0$, we must in priniciple resort to \eqref{eq:ffg} which is valid for any $\Delta\geq 0$. In practice however, this is equal to the analytic continuation of \eqref{eq:fAction} in $\Delta$. We will see that in the extremal functionals constructed using our prescription, $\Delta_0$ will coincide with the value of the gap we are maximizing or the scaling dimension of the operator whose OPE coefficient we are maximizing.

Since both $z^{2\Df}g(1-1/z)$ and the kernel of the second integral transform in \eqref{eq:intT} are real for $z\in(1,\infty)$, also $\mathfrak{g}(\Delta)$ is real. In all cases of interest, $g(z)$ will in fact be positive for $0<z<1$. Since the kernel is also positive, $\mathfrak{g}(\Delta)$ is positive too whenever $\Delta$ is sufficiently large so that the integral in \eqref{eq:intT} converges. On the other hand, $\mathfrak{f}(\Delta)$ may not be real, so let us write
\be
\mathfrak{f}(\Delta) = r(\Delta)e^{- i\pi\delta(\Delta)}\;\textrm{ with }\;r(\Delta)\in\mathbb{R}_{\geq0}\textrm{ and }\delta(\Delta)\in\mathbb{R}\,.
\ee
The functional action becomes
\be
\omega(\Delta) =  \mathfrak{g}(\Delta) - \cos\{\pi[\Delta-2\Df+\delta(\Delta)]\}r(\Delta)\,.
\ee
If $\mathfrak{g}(\Delta)$, $r(\Delta)$ and $\delta(\Delta)$ are slowly varying, the local minima of $\omega(\Delta)$ are near the maxima of the cosine, namely at
\be
\Delta_n \approx 2\Df + 2n - \delta(\Delta_n)\,,\quad n=0,1,\ldots
\label{eq:spec}
\ee
Moreover, provided $\mathfrak{g}(\Delta)\geq r(\Delta)$, the functional is non-negative. If $\omega$ is an extremal functional, the local minima become double zeros at the locations of the extremal spectrum, in which case we obtain
\be
\mathfrak{g}(\Delta) \approx r(\Delta)\,.
\label{eq:match}
\ee
In the following two sections, we will analyze examples where the approximate equalities in \eqref{eq:spec} and \eqref{eq:match} become exact, namely the theory of a generalized free fermion, and the scaling limit corresponding to gapped theories in large $AdS_2$.

\section{Functionals for the generalized free fermion}
\label{sec:free}

\subsection{General remarks}
Reference \cite{Gaiotto:2013nva} provided numerical evidence that the unitary 1D four-point function of identical operators with maximal gap above identity arises in the theory of the generalized free fermion, or equivalently the massive free fermion in $AdS_2$. The optimal four-point function in the region $0<z<1$ reads
\be
\mathcal{G}(z) = 1 +\left(\frac{z}{1-z}\right)^{2\Df} \!\!- z^{2\Df}\,.
\ee
The spectrum exchanged in the OPE consists of the identity and operators with dimension
\be
\Delta_n = 2\Df + 2n + 1\,,\quad n=0,1,\ldots
\ee
The bootstrap sum rule \eqref{eq:bootstrap1} reads
\be
F_0^{\Df}(z) + \sum\limits_{n=0}^{\infty}c_{n}^2 F_{\Delta_n}^{\Df}(z) = 0\,,
\label{eq:crossingFree}
\ee
where
\be
c_n^2 = \frac{2 (2 \Df)_{2 n+1}^2}{(2 n+1)! (4 \Df+2 n)_{2 n+1}}\,.
\ee
The extremality of this solution to crossing was demonstrated for $\Df\in\mathbb{N}/2$ in \cite{Mazac:2016qev} by analytically constructing the corresponding extremal functionals. We will now use the formalism of the previous section to construct the extremal functionals for any $\Df>0$ and thus prove the extremality in general.

The simplest way to ensure that the functional \eqref{eq:fActionT} has local minima at $\Delta_n$ is to require that $f(z)$ is real and negative for $z\in(1,\infty)$. If that is the case, $\delta(\Delta) = -1$ for all $\Delta$ for which the defining integral of $\mathfrak{f}(\Delta)$ converges. In order to make the minima into double zeros, we also require $\mathfrak{g}(\Delta) = -\mathfrak{f}(\Delta)$, which is equivalent to
\be
g(z) = - (1-z)^{2\Df - 2}f\!\left(\mbox{$\frac{1}{1-z}$}\right)\,.
\label{eq:gFree}
\ee
Under these constraints, the action of the functional \eqref{eq:fAction} reads
\be
\omega(\Delta) = 2\cos^2\left[\frac{\pi}{2}(\Delta-2\Df)\right]\int\limits_{1}^{\infty}\!\!dz
[-f(z)]\frac{\widehat{G}_{\Delta}(1-z)}{(z-1)^{2\Df}}\quad\textrm{for } \Delta>\Delta_0\,,
\label{eq:fActionFree}
\ee
where $\Delta_0$ is such that the integral converges. Very importantly, we must remember that $f(z)$ and $g(z)$ are also tied by the gluing condition \eqref{eq:glue}. Using \eqref{eq:gFree}, the gluing condition gives us the fundamental relation satisfied by $f(z)$ for the generalized free fermion
\be
\boxed{
\vphantom{\bigg|} 
\mathrm{Re}[f(z)]=z^{2\Df-2} f\!\left(\mbox{$\frac{1}{z}$}\right)+(1-z)^{2\Df-2} f\!\left(\mbox{$\frac{1}{1-z}$}\right)\quad\textrm{for } z\in (0,1)}\,.
\label{eq:fundamentalfree}
\ee
Recall that $f(z)$ in the lower half-plane is defined by $f(z) = f(1-z)$. Since $f(z)$ is real for $z\in(1,\infty)$, it follows that $f(z)$ is analytic in $\mathbb C\backslash[0,1]$ and therefore has a series expansion around $z=\infty$,
\be
f(z) = \sum\limits_{j=0}^{\infty} a_j w^{-j-1}\,,
\ee
where $w = z(z-1)$. We will see shortly that analyticity away from $z\in[0,1]$, together with the fundamental relation written above, allow us to fix $f(z)$.

The first step is to understand the boundary conditions. Notice that the integral in \eqref{eq:fActionFree} diverges for sufficiently small $\Delta$. The precise value of $\Delta$ where this happens depends on the behaviour of $f(z)$ as $z\rightarrow 1$. However, we can be sure that the true value of the functional $\omega(\Delta)$ is finite for all $\Delta\geq 0$, as manifested by \eqref{eq:fh} or \eqref{eq:ffg}. Therefore, the integral can only diverge at $\Delta\geq 0$ if the divergence is cancelled by a zero of the prefactor. The prefactor has double zeros at $\Delta_n = 2\Df+2n+1$, which means $f(z)$ is restricted to behave as
\be
f(z)\stackrel{z\to 1^+}{=} \frac{a \log(z-1) + b}{(z-1)^{2n+2}}+\textrm{subleading}
\quad\textrm{for some }n\in\mathbb{Z}\,.
\label{eq:z1behaviour}
\ee
When $a$ vanishes, the singularity of the integral is a simple pole in $\Delta$ at $\Delta_n$, combining with the double zero of the prefactor to give a simple zero of $\omega(\Delta)$. For $a$ non-zero, the singularity is a double pole at $\Delta_n$, leading to a finite nonzero $\omega(\Delta_n)$. It follows from the analytic properties of $f(z)$ and \eqref{eq:fundamentalfree} that necessarily $n\geq 0$, see Appendix \ref{app:boundN} for details. Since for $f(z)$ negative in the $z>1$ region we are guaranteed positivity of the functional beyond $\Delta_n$, and we would like the functionals to be positive in as wide as region as possible, we will set $n=0$.\footnote{The full set of solutions with arbitrary $n$ will be explored in detail in the next paper of this series~\cite{Mazac:2018ycv}.} 

We see that for fixed $\Df$ and $n=0$ there are essentially two distinct functionals, labeled by their behaviour near $z=1$. As we will see shortly, they correspond to gap maximization and OPE maximization functionals. In the first case we set $a=0$, call the resulting functional the \emph{normal functional} and denote it by $\beta$. The normal functional vanishes on all $\Delta_n$, with $\Delta_0$ being a simple zero and $\Delta_n$ for $n>0$ being double zeros. Therefore, by virtue of the crossing equation \eqref{eq:crossingFree} and the swapping condition, it also vanishes at $\Delta = 0$. We will normalize $\beta$ by requiring
\be
\partial_{\Delta}\beta(\Delta_0) = 1\quad\Leftrightarrow\quad
f(z)\stackrel{z\to 1}{\sim}-\frac{2}{\pi^2(z-1)^2}\,.
\ee
Similarly, we can construct the \emph{logarithmic functional}, for which $a\neq 0$, and which we will denote by $\alpha$. The dimensions $\Delta_n$ for $n\geq 1$ are again double zeros of $\alpha(\Delta)$, but now $\alpha(\Delta_0)\neq 0$. We will normalize $\alpha$ so that
\be
\alpha(\Delta_0) = 1\quad\Leftrightarrow\quad
f(z)\stackrel{z\to 1}{\sim}\frac{2\log(z-1)}{\pi^2(z-1)^2}\,.
\label{eq:z1Log}
\ee
We have a freedom to add a multiple of the normal functional to the logarithmic functional since this will not modify the logarithmic asymptotic behaviour as $z\rightarrow1$. We will fix this ambiguity by requiring $\partial_{\Delta}\alpha(\Delta_0) = 0$, which is equivalent to the absence of the $(z-1)^{-2}$ term in the expansion of $f(z)$ as $z\rightarrow1$. The swapping condition together with the crossing equation \eqref{eq:crossingFree} imply that the OPE coefficient squared $c^2_0$ can be read off from the action of $\alpha$ on the identity,
\be
\alpha(0) = -c_{0}^2\,.
\ee

Given these functionals, let us now discuss how they may be used to derive bounds on possible solutions to crossing. We begin with gap maximization. Consider the functional
\be
\omega_{\textrm{gapmax}} = \beta - \epsilon\,\alpha
\ee
for $\epsilon>0$. The action on identity is positive
\be
\omega_{\textrm{gapmax}}(0) = \epsilon\, c_{0}^2 > 0\,.
\ee
For $\epsilon\ll 1$, the simple zero of $\beta(\Delta)$ at $\Delta = \Delta_0$ gives rise to a simple zero of $\omega_{\textrm{gapmax}}(\Delta)$ at
\be
\tilde{\Delta}_0 = \Delta_0 + \epsilon + O(\epsilon^2)\,.
\ee
Moreover, we will be able to show from the explicit form of $\alpha$ and $\beta$ given in the following section that for sufficiently small $\epsilon$, $\omega_{\textrm{gapmax}}(\Delta)$ is non-negative to the right of $\tilde{\Delta}_0$
\be
\omega_{\textrm{gapmax}}(\Delta) \geq 0\quad\textrm{for }\Delta\geq\tilde{\Delta}_0\,.
\ee
Following the discussion in section \ref{sec:optbounds}, this functional shows that there must be at least one primary operator with $0<\Delta<\tilde{\Delta}_0$. Taking the $\epsilon \rightarrow 0$ limit demonstrates that the optimal gap above identity is $\Delta_0=2\Df+1$, with corresponding extremal functional given by $\beta$, and with associated extremal solution to crossing the generalized free fermion.

Similarly, we can use $\alpha$ and $\beta$ to derive upper bounds on OPE coefficients. Consider
\be
\omega^{t}_{\textrm{opemax}}= 
\alpha+t\,\beta
\label{eq:omOPEmax}
\ee
for $t\in\mathbb{R}$.
We have
\ba
&\omega^{t}_{\textrm{opemax}}(\Delta)\geq 0\quad \textrm{for}\quad \Delta\in S(t)\qquad \\
&\omega^{t}_{\textrm{opemax}}(\Delta_0)=1, \qquad \,\,\omega^{t}_{\textrm{opemax}}(0)=-c_0^2\,, \qquad 
\ea
where the first line should be viewed as a definition of the set $S(t)$. As reviewed in section \ref{sec:review}, for each $t$, this functional provides an upper bound on the OPE coefficient of an operator of dimension $\Delta_0$ provided all other operators are within $S(t)$. The precise form of $S(t)$ depends on the details of $\alpha(\Delta)$ and $\beta(\Delta)$ and we will comment on it more in the next section armed with the explicit formulas given there. The upper bound on the OPE coefficient is independent of $t$ and equal to $c_0^2$. The bound is optimal since again the GFF solution saturates it.\footnote{We should note that $\omega^{t}_{\textrm{opemax}}$ provides an upper bound for the OPE coefficient of an operator of arbitrary dimension $\Delta^*$, as long as $\omega^{t}_{\textrm{opemax}}(\Delta^*)>0$, but this bound will only be optimal for $\Delta^*=\Delta_0$.}

Before we move on, let us summarize our findings. The claim is that extremal functionals associated to the generalized free fermion are obtained from a function $f(z)$ with certain properties. It should be analytic in $\mathbb{C}\backslash[0,1]$ and satisfy $f(z) = f(1-z)$. The boundary conditions are such that $f(z)$ must decay at least as $z^{-2}$ as $z\to \infty$, and $f(z)= [a \log(z-1)+b](z-1)^{-2}+\textrm{subleading}$ as $z\rightarrow 1^+$. Crucially, $f(z)$ is constrained by the fundamental relation \eqref{eq:fundamentalfree}, which is the only place where the external dimension $\Df$ enters the problem. Finally, we must check whether appropriate choices of $a$ and $b$ can be made such that $f(z)\leq 0$ for $z\in(1,\infty)$, thus guaranteeing the positivity of the functional action above $2\Df + 1$.

\subsection{Constructing the normal functionals}
\label{sec:construct}
We will proceed by finding $f(z)$ corresponding to the normal functionals $\beta$ for any $\Df>0$. 
Let us first find $f(z)$ for $\Df \in \mathbb{N}-1/2$. From the results of \cite{Mazac:2016qev}, we can read off $f(z)$ for $\Df=1/2,3/2$ and 5/2:
\ba
\Df=1/2\,:\quad f(z)&=-\frac{5w+2}{\pi^2 w^2}\\
\Df=3/2\,:\quad f(z)&=-\frac{7w+2}{\pi^2 w^2}\\
\Df=5/2\,:\quad f(z)&=\frac{\left[6w^2(w-2)(2z-1)\log(\mbox{$\frac{z-1}z$})+(w+1)(12w^2-35w-10)\right]}{5\pi^2 w^2}\,,
\ea
where $w = z(z-1)$. All the required properties of $f(z)$ hold in these examples. This suggests the following general ansatz for $f(z)$ when $\Df \in \mathbb{N}-1/2$:
\be
f(z)=\frac{1}{w^{2}}\left[\left(\sum_{k=0}^{K} a_k w^k\right)(2z-1) \log(\mbox{$\frac{z-1}z$})+\sum_{k=0}^{K} b_k w^k\right],
\label{eq:fAnsatz}
\ee
for some integer $K$, with $a_k$ and $b_k$ to be determined. The fundamental relation \eqref{eq:fundamentalfree}, together with the asymptotic conditions as $z\rightarrow 1^+$ and $z\rightarrow \infty$ lead to a unique solution for $a_k$, $b_k$. The solution for general $\Df \in \mathbb{N}-1/2$ is not particularly enlightening in the $z$-variable. However, it becomes much more elegant after performing a version of the Mellin transform. Let us define the following Mellin-like transform of $f(z)$
\be
M(s) = -\frac{1}{2\cos(\pi s)}\int\limits_{0}^1\!\!dz \,[z(1-z)]^s\mathrm{Re}[f(z)]\,,
\label{eq:mellinDef}
\ee
where the prefactor is inserted for later convenience. As we discuss in Appendix \ref{app:mellin}, the only poles of $M(s)$ for $\mathrm{Re}(s)>0$ are a simple pole at $s=1$ and the poles of the prefactor at $s=1/2 + n$, $n=0,1,\ldots$. We also show there that the transform can be inverted to give
\be
f(z) = \frac{2z-1}{z(z-1)}\!\int\limits_{\Gamma}\!\!\frac{ds}{2\pi i}\,[z(z-1)]^{-s}M(s)\,,
\label{eq:invMellin}
\ee
where the contour $\Gamma$ goes from $s=-i\infty$ to $s=i\infty$ to the left of the poles at $s=1/2 + n$, $n=0,1,\ldots$, but to the right of the pole at $s=1$.

Computing $M(s)$ for $\Df\in\mathbb{N}-1/2$ from \eqref{eq:fAnsatz} with $a_k,b_k$ fixed by the constraints on $f(z)$, we found the general formula
\be
\boxed{
M_{\beta}(s) = 
\frac{(2 \Df+3 s) \Gamma \left(\Df+\frac{3}{2}\right) \Gamma \left(\frac{1}{2}-s\right) \Gamma (s-1) \Gamma (s+1) \Gamma (2 \Df+s+1)}
{4^{\Df+s}\pi ^2 \Gamma (\Df+1) \Gamma \left(\Df+s+\frac{1}{2}\right) \Gamma \left(\Df+s+\frac{3}{2}\right)}}\,.
\label{eq:mellinM}
\ee
Note that $M_\beta(s)$ is an analytic function of $\Df$ for $\Df\geq 0$. It is therefore natural to expect that $M_\beta(s)$ is the transform of the correct $f(z)$ not just for $\Df\in\mathbb{N}-1/2$ but for general $\Df > 0$. We can now use the Mellin inversion formula \eqref{eq:invMellin} to find $f_\beta(z)$ in general:
\ba
f_{\beta}(z)=
-\kappa(\Df)
&\frac{2z-1}{w^{3/2}}\left[\, _3\tilde{F}_2\left(-\frac{1}{2},\frac{3}{2},2
   \Df+\frac{3}{2};\Df+1,\Df+2;-\frac{1}{4 w}\right)+\right.\\
	&\,\;\;\left.+\frac{9}{16 w} \,
   _3\tilde{F}_2\left(\frac{1}{2},\frac{5}{2},2 \Df+\frac{5}{2};\Df+2,\Df+3;-\frac{1}{4w}\right)\right],
\label{eq:allaf}
\ea
where ${}_3\tilde F_2$ stands for the regularized hypergeometric function, $w=z(z-1)$ and the normalization factor reads
\be
\kappa(\Df) = \frac{\Gamma(4\Df+4)}{2^{8\Df+5}\Gamma(\Df+1)^2}\,.
\ee
Although we have only derived \eqref{eq:allaf} for certain $\Df \in \mathbb{N}-1/2$, we can now check whether it satisfies all the requirements for any $\Df>0$. Note that the prefactor $(2z-1)w^{-3/2}$ is symmetric\footnote{To be precise, the analytic continuation of the function that for $z>1$ is given by $(2z-1)[z(z-1)]^{-3/2}$, is symmetric under $z\to 1-z$.} under $z\rightarrow1-z$, so that $f_\beta(z) = f_\beta(1-z)$. We can check exactly that the asymptotic behaviour $f_\beta(z)\stackrel{z\rightarrow 1}{\sim}-2\pi^{-2}(z-1)^{-2}$ holds for all $\Df\geq 0$. We have also checked numerically to high accuracy that the fundamental relation \eqref{eq:fundamentalfree} is satisfied. The validity of the fundamental relation for general $\Df$ can in fact be proven analytically using a certain third-order ODE satisfied by $f_\beta(z)$, as explained in Appendix \ref{app:difeq}. In order to test the condition $f_\beta(z)\leq0$ for $z\in(1,\infty)$, we can study the series expansion of $f_\beta(z)$ around $z=\infty$. Since the only non-analyticity of $f_\beta(z)$ is the branch cut at $z\in[0,1]$, this expansion must be convergent for all $z\in(1,\infty)$. The first few terms of the expansion read 
%
\ba
f_\beta(z) \stackrel{z\rightarrow\infty}{\sim} -\frac{2\kappa(\Df)}{\Gamma(\Df+1)\Gamma(\Df+2)}&\left[
\frac{1}{z^2}
+\frac{1}{z^3}
+\frac{3 \left(2 \Delta _{\phi }+3\right) \left(6 \Delta _{\phi }+11\right)}{32 \left(\Delta _{\phi }+1\right) \left(\Delta _{\phi }+2\right)z^4}
+O(z^{-5})
\right]\,.
\ea

We can see that $f_{\beta}(z)$ enjoys the correct supression at $z=\infty$ and that the terms in the expansion are negative for $\Df > 0$. We verified the negativity of the coefficients up to $O(z^{-150})$, providing strong evidence that indeed $f(z)<0$ for $z\in(1,\infty)$.\footnote{For $z>2$ we can prove $f_{\beta}(z)$ is manifestly negative by means of the transformation:
\be
_3F_2\left(a,1-a,b;\frac{b+a+1}2,1+\frac{b-a}2;y\right)=(1-4y)^{-b}\, _3F_2\left(\frac b3,\frac{b+1}3,\frac{b+2}3;\frac{b+a+1}2,1+\frac{b-a}2,-\frac{27 y}{(1-4y)^3}\right),\nonumber
\ee
valid for $-1/8<y<1/4$. Unfortunately we are not aware of a similar transformation for $y<-1/8$.} 
This completes the necessary checks that \eqref{eq:allaf} defines the normal functional for the generalized free fermion with all the required properties.

It is interesting to note that while the general formula for $f_{\beta}(z)$ is rather complicated, it simplifies greatly in the $\Df\rightarrow\infty$ limit
\be
f_{\beta}(z) \stackrel{\Df\rightarrow\infty}{\sim} - \sqrt{\frac{2\Df}{\pi^3}}\frac{2z-1}{[z(z-1)]^{3/2}}\,.
\label{eq:fFreeInf}
\ee
The double poles at $z=0,1$ are subleading in this limit.

\begin{figure}[ht!]%
\begin{center}
\includegraphics[width=12cm]{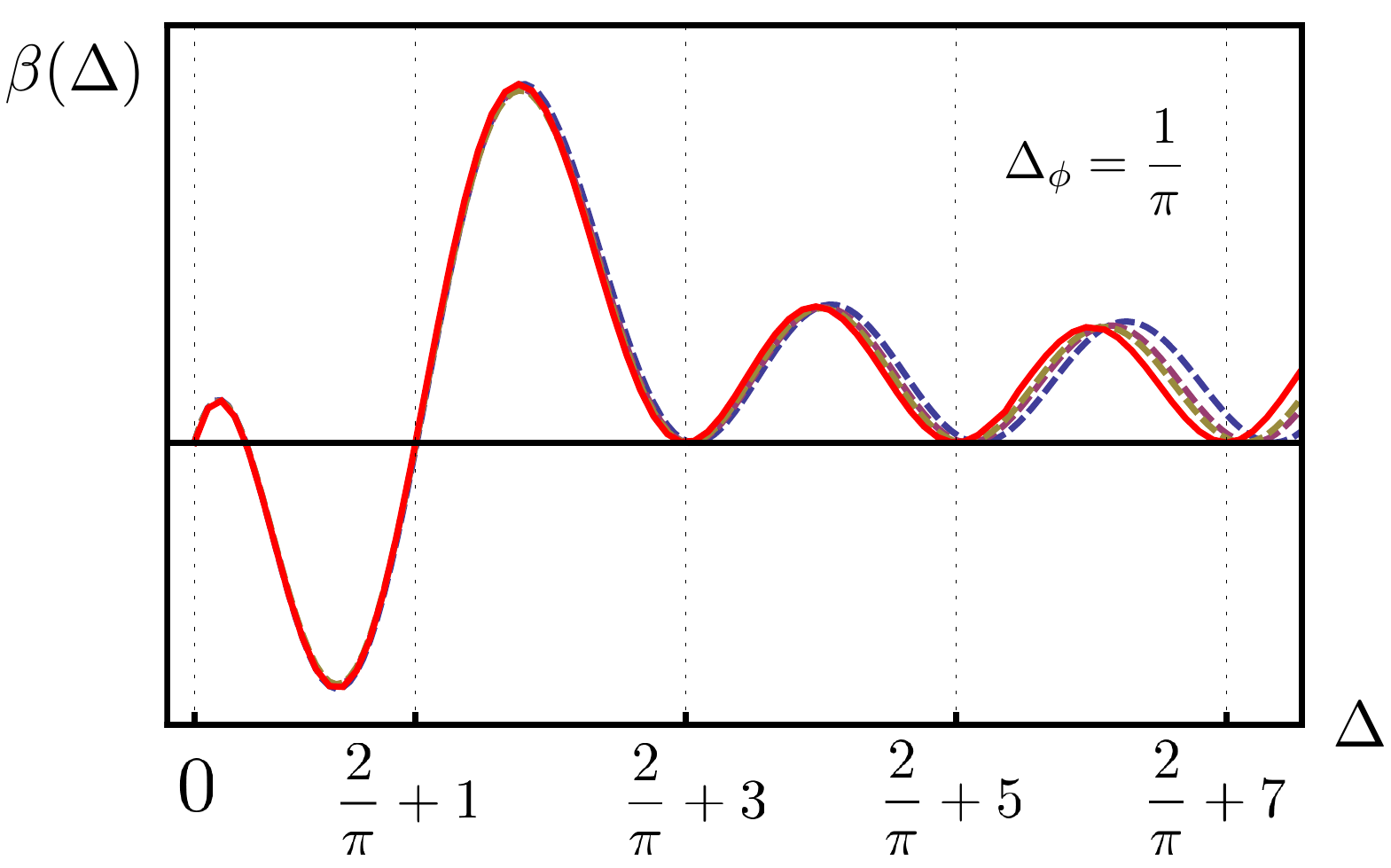}%
\caption{Gap maximization functional at $\Df=1/\pi$. The three dashed curves are numerical results obtained using {\tt JuliBootS} and the flow method~\cite{Paulos:2014vya,El-Showk2016} with $N=104,184$ and 264 derivatives. As the number of components is increased, the functional action converges to the red curve, with a simple zero at $\Delta=2\pi^{-1}+1$ and double zeros for $\Delta=2\pi^{-1}+2n+1$ with $n\geq 1$. The red curve in turn was obtained by acting with the analytic normal functional \eqref{eq:allaf} on the $F_\Delta^{\Df}$ vectors as in \eqref{eq:ffg}.} %
\label{fig:gfffunctional}%
\end{center}
\end{figure}

Last but not least, we can compare the action of our analytic functional with the results of the numerical bootstrap. Figure \ref{fig:gfffunctional} provides this comparison for the transcendental value\footnote{Because why not?} $\Df = 1/\pi$, showing excellent agreement.

\subsection{Constructing the logarithmic functionals}
\label{sec:constructLog}
In order to find the logarithmic functionals $\alpha$, characterized by \eqref{eq:z1Log} and the condition $\partial_{\Delta}\alpha(\Delta_0) = 0$, we can repeat the steps we used for the normal functionals. The ansatz \eqref{eq:fAnsatz} still works for $\Df\in\mathbb{N}-1/2$, and we get a logarithmic functional precisely when $a_0 \neq 0$, $b_0=0$. The fundamental relation together with asymptotic conditions at $z=1,\infty$ again fix $a_k$ and $b_k$ uniquely. Computing the Mellin transform \eqref{eq:mellinDef}, we find it only differs slightly from the Mellin transform of the normal functionals
\be
M_{\alpha}(s) = 
\left[\frac{1}{s-1}+\frac{1}{s}-\frac{1}{2s+2 \Delta _{\phi }+1}+\frac{3H\!\left(\Df+\frac{1}{2}\right)}{2}-\frac{H(\Delta _{\phi})}{2}-\log (2)\right]M_{\beta}(s)\,,\label{eq:logmellin}
\ee
where $H(s)$ is the harmonic number. $f_{\alpha}(z)$ is given by the Mellin inversion formula~\eqref{eq:invMellin}. We will find it convenient to separate the last three terms in the square bracket as follows
\be
f_{\alpha}(z) = \tilde{f}_{\alpha}(z) +\left[\frac{3}{2}H\!\left(\Df+\frac{1}{2}\right)-\frac{1}{2}H(\Delta _{\phi})-\log (2)\right] f_{\beta}(z)\,,
\label{eq:fAlphaSplit}
\ee
where
\ba
\tilde{f}_{\alpha}(z)=
\kappa(\Df)\frac{2(z-2)(z+1)}{(2z-1)w^{3/2}}
&\left[
{}_3\tilde{F}_2\left(-\frac{1}{2},-\frac{1}{2},2\Df +\frac{3}{2};\Df +2,\Df +2;-\frac{1}{4 w}\right)+\right.\\
+\frac{(2 \Df +3) (2 \Df +5)}{16 w} &{}_3\tilde{F}_2\left(\frac{1}{2},\frac{1}{2},2 \Df +\frac{5}{2};\Df +3,\Df +3;-\frac{1}{4 w}\right)-\\
-\frac{3 (4 \Df +5)}{256 w^2}&\left.{}_3\tilde{F}_2\left(\frac{3}{2},\frac{3}{2},2 \Df +\frac{7}{2};\Df +4,\Df +4;-\frac{1}{4 w}\right)\right].
\label{eq:fAlphaTilde}
\ea
Setting the question of positivity aside for now, we checked that the functional specified by $f_{\alpha}(z)$ has all the other necessary properties. In particular, the second term in \eqref{eq:fAlphaSplit} ensures that $\partial_{\Delta}\alpha(\Delta_0) = 0$.

We would now like to understand better the general OPE maximization functional
\be
\omega^{t}_{\textrm{opemax}}= 
\tilde{\alpha}+t\,\beta\,,
\label{eq:opeMax22}
\ee
where $\tilde{\alpha}$ is the functional arising from $\tilde{f}_{\alpha}$, and we made a different choice for the origin $t=0$ with respect to \eqref{eq:omOPEmax}. The first thing to notice is that $\tilde{f}_{\alpha}(z)$ cannot be negative for all $z>1$ because of the prefactor $z-2$. In fact, one can check that the sum of generalized hypergeometrics in the square bracket is positive for $z>1$ so that $\tilde{f}_{\alpha}(z)<0$ for $1<z<2$ and $\tilde{f}_{\alpha}(z)>0$ for $z>2$. $\tilde{\alpha}(\Delta)$ with $\Delta\gg 1$ probes $\tilde{f}_{\alpha}(z)$ for large values of $z$ and thus $\tilde{\alpha}(\Delta)$ has a negative region for sufficiently large $\Delta$.

On the other hand, expanding $\tilde{f}_\alpha(z)$ at large $z$, we find that only the coefficient of $z^{-2}+z^{-3}$ is positive and all the higher ones are negative. We can make the coefficient of $z^{-2}+z^{-3}$ in $\tilde{f}_{\alpha}(z)+t f_\beta(z)$ also negative by requiring
\be
t\geq t_{c} \equiv \frac{1}{2(\Df+1)}\,.
\ee
It follows that $\omega^{t}_{\textrm{opemax}}(\Delta)$ is nonnegative for all $\Delta\geq2\Df + 1$ if and only if $t\geq t_c$. When the inequality is saturated, the resulting $f(z)$ has a $z^{-4}$ falloff as $z\rightarrow\infty$.

Therefore, provided $t\geq t_c$, $\omega^{t}_{\textrm{opemax}}(\Delta)$ is the extremal functional for the problem of maximizing the OPE coefficient of an operator with $\Delta = 2\Df + 1$ such that all other primaries in the OPE have $\Delta\geq\Delta^*$, where $\Delta^*$ is the largest simple zero of $\omega^{t}_{\textrm{opemax}}(\Delta)$. In order for $\omega^{t}_{\textrm{opemax}}(\Delta)$ to be maximally constraining, we should choose $t$ such that $\Delta^*$ is minimal. It is possible to convince oneself that this happens precisely for the minimal $t$, i.e. $t=t_c$. We will denote the largest simple zero of $\omega^{t_c}_{\textrm{opemax}}(\Delta)$ by $\Delta^{*}_{c}$.

In the end, we have arrived at the following non-trivial prediction. Consider the upper bound on the OPE coefficient at $\Delta=2\Df + 1$ as a function of the lower bound $\Delta^*$ we impose on the scaling dimension of all other operators in the OPE. As long as $\Delta^*\geq\Delta^*_c$, the upper bound is given by the constant value $c_0^2$ in the generalized free fermion theory. For $\Delta^*<\Delta^*_c$ however, the generalized free fermion stops being the extremal solution and the character of the bound must change. The value of $\Delta^*_c$ can be derived from our explicit formulas in principle to any precision. For example, for $a=1/2$ we find $\Delta^*_c=0.6770671915683\ldots$. This completes an analytic explanation of a toy mechanism leading to a sharp feature in a conformal bootstrap bound at a nontrivial location.

\begin{figure}[ht!]%
\begin{center}
\includegraphics[width=12.5cm]{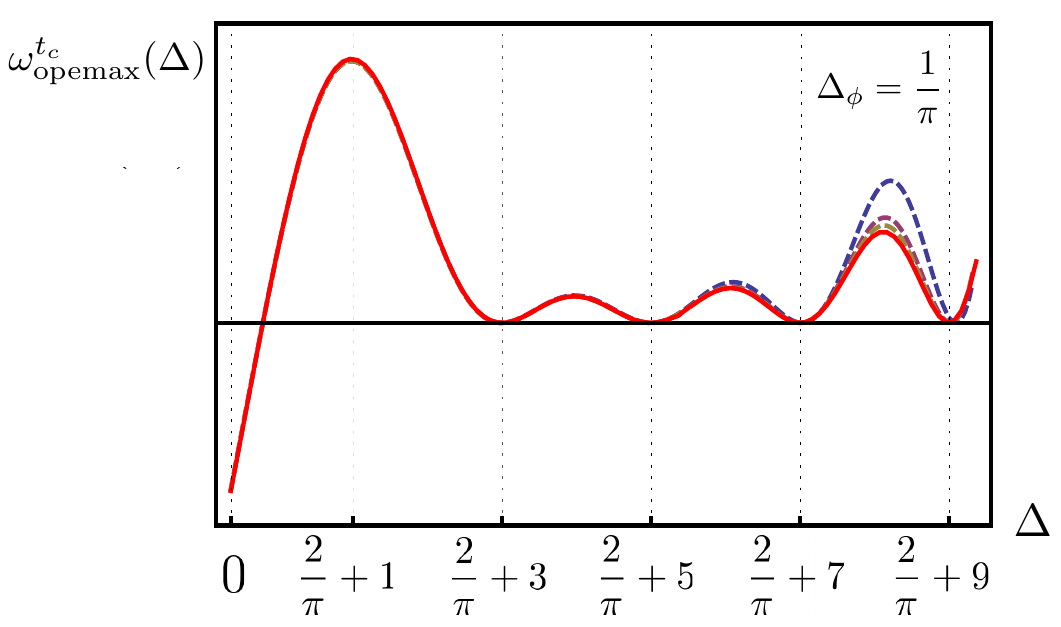}%
\caption{OPE maximization functional at $\Df=1/\pi$. The three dashed curves which are almost entirely overlapping are the numerical results obtained using {\tt JuliBootS} \cite{Paulos:2014vya,El-Showk2016} with $N=104,184$ and 264 derivatives. As the number of components is increased, the functional action rapidly converges to the red curve, with double zeros for $\Delta=2\pi^{-1}+2n+1$ with $n\geq 1$. This curve in turn was obtained by acting with the combination of logarithmic and normal functionals that decays as $\sim z^{-4}$ as $z\rightarrow\infty$.}%
\label{fig:gffopefunctional}%
\end{center}
\end{figure}

Numerical bootstrap confirms our prediction. Specifically, it appears that the ope maximization problem is unbounded for $0<\Delta^*<\Delta^*_c$. On the other hand, for any $\Delta^*\geq\Delta^*_c$, the numerical bootstrap algorithm tends to reconstruct precisely the functional with $z^{-4}$ fall-off. Figure \ref{fig:gffopefunctional} contains a comparison of the numerical functionals and the analytic functional \eqref{eq:opeMax22} with $t=t_c$, showing perfect agreement.

\section{OPE maximization at large $\Delta$} \label{sec:largeR}
\subsection{The problem}
In this section, we will apply our formalism to the following interesting bootstrap problem. We consider unitary solutions to crossing \eqref{eq:bootstrap1} such that there is a single primary operator $\mathcal{O}_b$ with scaling dimension $\Delta_b$ in the region $0<\Delta_b<2\Df$. We do not impose any constraints on the spectrum above $2\Df$. We ask what is the maximal value of the OPE coefficient $c_b^2\equiv c_{\phi\phi\mathcal{O}_b}^2$ among such solutions to crossing. As explained in section \ref{sec:optbounds}, an upper bound on $c_b^2$ can be obtained from a functional $\omega$ satisfying
\be
\omega(\Delta_b)>0\quad\textrm{and}\quad\omega(\Delta)\geq0\quad\textrm{for }\Delta\geq2\Df\,.
\label{eq:opeConstraints}
\ee
We then necessarily have $\omega(0)<0$ and the upper bound reads
\be
c_b^2 \leq -\frac{\omega(0)}{\omega(\Delta_b)}\,.
\label{eq:opeBound}
\ee
The optimal (lowest) upper bound is obtained by minimizing this ratio by scanning over all functionals subject to constraints \eqref{eq:opeConstraints}. In the optimal solution to crossing, the inequality \eqref{eq:opeBound} is saturated, with $\omega$ being the extremal functional. The spectrum of the optimal solution consists of $\mathcal{O}_b$ and (possibly a subset of) the zeros of $\omega(\Delta)$ for $\Delta\geq 2\Df$.

The analysis of the general problem is beyond the scope of this work. Instead, we will solve the optimization problem analytically at the leading order in the limit where
\be
\Df,\Delta_b\rightarrow\infty\quad\textrm{with}\quad m \equiv \frac{\Delta_b}{\Df}\in(0,2)\quad\textrm{fixed.}
\label{eq:largeR}
\ee
We will find the extremal functional in this limit and use it to derive the optimal bound as well as the spectrum of the optimal solution to crossing. We will see the spectrum corresponds to an interacting 1D CFT parametrized by $m\in(0,2)$. The key observation allowing us to solve the problem is that the action of the functional can be computed using a saddle-point approximation in this limit.

This limit has already been considered in \cite{Paulos:2016fap} using the numerical conformal bootstrap. As discussed in that article, an important class of solutions to crossing arises from placing any massive (unitary and UV-complete) 2D QFT into $AdS_2$. Scaling dimensions of primary operators are then proportional to the radius of $AdS$. The limit corresponds to taking the $AdS$ radius large while keeping the flat-space masses and couplings fixed.  Our set-up then corresponds to a scattering process $\phi\phi\rightarrow\phi\phi$, where the $\phi$ particle is the lightest particle in the theory which has a single flat-space bound state of mass $m$ measured in the units of mass of the $\phi$-particle.\footnote{In reference \cite{Paulos:2016fap}, $m_1, m_b$ are used for the masses of the external particle and its bound state respectively. Therefore our $m$ is $m_b/m_1$ in that reference.}  An upper bound on $c_b^2$ corresponds to an upper bound on the non-perturbative coupling between two $\phi$-particles and the bound state. Physically, such a bound should exist since increasing the coupling increases the force mediated by the bound state, eventually causing new bound states to appear, thus invalidating our original assumption on the spectrum. We will see that our extremal solutions to crossing correspond to integrable field theories placed in large $AdS_2$, as observed numerically in \cite{Paulos:2016fap}.\footnote{For $\phi$ to be the lightest particle of the theory, we should have $1<m<2$. While we will solve the CFT problem for all $0<m<2$, we do not know if the extremal solution has a physical interpretation for $0<m<\sqrt{2}$.} We will sometimes refer to \eqref{eq:largeR} as the \emph{large-radius} limit although our results are completely general and do not rely on an underlying $AdS$ description.

\subsection{The solution}\label{ssec:solution}
We will work in the subspace of bootstrap functionals given by \eqref{eq:ffg} with $f(z)$ and $g(z)$ satisfying constraints discussed in section \ref{ssec:funcsGen}. We will find the extremal functional in this subspace by optimizing over $f(z)$ and $g(z)$. A priori, there is no guarantee that the true extremal functional for the full problem lies in this subspace. However, we will be able to prove the optimality of our functional. We will do so by exhibiting a physical solution to crossing (asymptotically in the large-radius limit) which saturates the bound arising from our functional, guaranteeing that no better functional exists. 

We set $\Delta_b = m \,\Df$ with $0<m<2$ and regard $\Df$ as a parameter. For any finite $\Df$, let us denote the extremal functional for the problem we want to solve by $\omega_{\Df}$, and the functions specifying it by $f_{\Df}(z)$ and $g_{\Df}(z)$. In order to make progress, we need to make a general assumption on the asymptotic behaviour of $f_{\Df}(z)$ and $g_{\Df}(z)$ as $\Df\rightarrow\infty$. We will take inspiration from the normal functionals for the free fermion, where we found\footnote{After renormalizing $\omega_{\Df}$ so that $f_{\Df}(z)$ has a finite limit.}
\ba
f^{\textrm{free}}_{\Df}(z) &\sim -\frac{2z-1}{[z(z-1)]^{3/2}}\\
g^{\textrm{free}}_{\Df}(z) &\sim (1-z)^{2\Df}\frac{(z+1)}{z^{3/2}}
\ea
as $\Df\rightarrow\infty$. Accordingly, we will assume that in general
\ba
f_{\Df}(z)&\sim f(z)\qquad\quad\qquad\textrm{for }\mathrm{Im}(z)>0\\
g_{\Df}(z)&\sim (1-z)^{2\Df}\tilde{g}(z)\,\,\;\textrm{for }z\in(0,1)
\label{eq:fglargeR}
\ea
in that limit, with $f(z)$ and $\tilde{g}(z)$ to be determined. Moreover, we will assume that the convergence is sufficiently uniform so that $f(z)$ satisfies the usual constraints on the $f$ weight-function. We will see that this asymptotic behaviour guarantees a well-behaved $\Df\rightarrow\infty$ limit of the functionals. Our goal in the remainder of this section is to find $f(z)$ and $\tilde{g}(z)$ corresponding to the asymptotic solution of our optimization problem. We will see that the optimal $f(z)$ is closely related to the flat-space S-matrix of a two-dimensional theory.

The optimization problem naturally splits into two parts: making sure that $\omega_{\Df}(\Delta)\geq0$ for $\Delta> 2\Df$ and minimizing the ratio $-\omega_{\Df}(0)/\omega_{\Df}(m\Df)$. We will start by analyzing the former condition. In the regime $\Delta>2\Df$, we can use the contour deformation of section \ref{ssec:funcSimp} to write the action of the functional as an integral over $z\in(1,\infty)$, i.e. \eqref{eq:fAction}, which we reproduce here for convenience\footnote{One reason to expect the contour deformation to be allowed for $\Delta>2\Df$ is by analogy with the free fermion functionals, where it is valid for $\Delta>2\Df + 1$.}
\be
\omega_{\Df}(\Delta) = \int\limits_{1}^{\infty}\!\!dz\!\left\{z^{2\Df - 2}\,g_{\Df}\!\left(\frac{z-1}{z}\right) - 
\mathrm{Re}\left[e^{-i\pi(\Delta-2\Df)}f_{\Df}(z)\right]\right\}\frac{\widehat{G}_{\Delta}(1-z)}{(z-1)^{2\Df}}
\,.
\label{eq:fAction5}
\ee
Demanding that this integral converges for $\Delta>2\Df$ implies $f_{\Df}(z)$ should not grow faster than $(z-1)^{-1+\epsilon}$ for some $\epsilon>0$ as $z\rightarrow1$, and that $g_{\Df}(z)$ should not grow faster than $z^{-1+\epsilon}$ for some $\epsilon>0$ as $z\rightarrow0$. We will assume the same is true for the limits $f(z)$ and $\tilde{g}(z)$.

We take the large-radius limit in the regime $\Delta > 2\Df$ by setting $\Delta = \sqrt{s}\,\Df$ with $s>4$ fixed and taking $\Df\rightarrow \infty$. In the language of QFT in large $AdS_2$, an operator of dimension $\sqrt{s}\Df$ exchanged in the $\phi\times\phi$ OPE corresponds to an intermediate state of center-of-mass energy $\sqrt{s}$ in the scattering process $\phi\phi\rightarrow\phi\phi$, measured in the units of the $\phi$-particle mass. In other words, $s$ is the standard flat-space Mandelstam variable $(p_1+p_2)^2$ in these units. In this limit, the integral \eqref{eq:fAction5} is dominated by a saddle point at
\be
z = \frac{s}{4}\,,
\ee
as we show in appendix \ref{app:saddleHigh}.\footnote{
The localization has a nice physical interpretation. Conformal blocks can be computed via geodesic Witten diagrams \cite{Hijano2016}, with a particle being exchanged between a pair of geodesics in $AdS_2$. In the limit under consideration the exchange takes place only between the nearest points on this pair. When we then integrate over $z$ the saddle point will occur where the geodesics intersect, so that they emanate from four boundary points and meet at an interaction point inside $AdS_2$. The immediate neighbourhood of the intersection point then has the kinematics of a flat-space scattering process with center-of-mass energy $\sqrt{s}$. We thank Shota Komatsu for discussions regarding this point.}
Note that the flat-space crossing tranformation of the Mandelstam variable
\be
s \mapsto 4-s
\ee
becomes the CFT crossing transformation $z\mapsto 1-z$. The final answer for the integral \eqref{eq:fAction5} in the large-radius limit reads
\be
\omega_{\Df}(\Delta)\sim \mu(\Df,s)\left\{\left(\mbox{$\frac{s}{4}$}\right)^{-2}\tilde{g}\left(\mbox{$\frac{s-4}{s}$}\right)
-\cos\!\left[\pi\!\left(\Delta-2\Df + \delta\!\left(\mbox{$\frac{s}{4}$}\right)\right)\right]\left|f\left(\mbox{$\frac{s}{4}$}\right)\right|
\right\}\,,
\label{eq:actionLargeR}
\ee
where $\mu(\Df,s)$ is a positive prefactor independent of $f(z)$ and $\tilde{g}(z)$ given in appendix \ref{app:saddleHigh}, and where we factored $f(z)$ on the branch cut into its phase and absolute value
\be
f(z+i 0^+) = |f(z)|e^{-i\pi\delta(z)}\,.
\ee
Note that in spite of the saddle-point localization, the integral does not approach a smooth function of $s$ as $\Df\rightarrow\infty$. This is because the second term in the curly bracket contains the term $\cos(\pi\Delta+\textrm{const.})$ and thus oscillates as function of $\Delta$ with period $\approx 2$. Therefore, the oscillations become infinitely fast as a function of $s$ in the $\Df\rightarrow\infty$ limit. The oscillations will be necessary to reproduce the spectrum of the extremal solution. Note that the phase of $f(s/4)$ controls the shift of the minima of the functional away from the free scalar values $2\Df + 2n$.

It is now straightforward to state the necessary and sufficient condition for the asymptotic non-negativity of the functional for any $s>4$. Since the oscillations become arbitrarily fast as a function of $s$ in our limit, the first term in the curly bracket must be greater than $|f(s/4)|$ for any $s>4$. Equivalently,
\be
\tilde{g}(z) \geq (1-z)^{-2}\left|f\left(\mbox{$\frac{1}{1-z}$}\right)\right|\quad\textrm{for }z\in(0,1)\,.
\ee

Having established the condition for the positivity of the functional, we can move on to minimizing $-\omega_{\Df}(0)/\omega_{\Df}(m\Df)$. We need to evaluate $\omega_{\Df}(m\Df)$ and $\omega_{\Df}(0)$ in the large-radius limit in terms of $f(z)$ and $\tilde{g}(z)$. Representation \eqref{eq:fAction5} is not available in this regime of $\Delta$, and we need to proceed differently. As we show in appendix \ref{app:saddleLow}, the computation of $\omega_{\Df}(m\Df)$ localizes to a saddle point at
\be
z_b = \frac{m^2}{4} \in (0,1)\,,
\ee
the result being 
\be
\omega_{\Df}(m\Df)\!\sim\!
\frac{1}{16}\sqrt{\frac{\pi }{\Df}}  m^{1/2}\!\left(2+m\right)^{3/2}\!\left(2-m\right)^{1/2}\!
\left[\!\frac{2^{2(m+2)}}{
\left(2-m\right)^{2-m}\left(2+m\right)^{2+m}}\!\right]^{\Df}
\!\!\!\!\!\!\!\mathrm{Im}\!\left[f(z_b)\right]
\label{eq:fActionOnBound}
\ee
as $\Df\rightarrow\infty$. In particular, $\omega_{\Df}(m\Df)$ asymptotically only depends on $f(z)$ and not on $\tilde{g}(z)$. Note that the factor multiplying $\mathrm{Im}\!\left[f(z_b)\right]$ on the rhs is always exponentially large (and positive) as $\Df\rightarrow\infty$ since
\be
\frac{2^{2(m+2)}}{
\left(2-m\right)^{2-m}\left(2+m\right)^{2+m}} > 1\quad\textrm{for }0<m<2\,.
\ee
Since $\omega_{\Df}(m\Df)$ should be positive for an OPE maximization functional, we need to impose
\be
\mathrm{Im}\!\left[f(z_b)\right]>0\,.
\ee
In order to evaluate the action on identity $\omega_{\Df}(0)$ in the large-radius limit, we can use a trick relying on the crossing-symmetry of the free-fermion four-point function. As we explain in more detail in appendix \ref{app:largeRidentity}, the limit simplifies greatly 
\be
\omega_{\Df}(0) \sim
-\int\limits_{0}^{1}\!\!dz\,\tilde{g}(z)\,,
\label{eq:fActionOnId}
\ee
i.e. the dependence on $f(z)$ drops out completely. The optimization problem can now be reformulated entirely in terms of the functions $f(z)$ and $\tilde{g}(z)$
\be
\textrm{minimize}\quad \frac{\int_{0}^{1}\!dz\,\tilde{g}(z)}{\mathrm{Im}[f(z_b)]}\quad\textrm{subject to}\quad
\tilde{g}(z) \geq (1-z)^{-2}\left|f\left(\mbox{$\frac{1}{1-z}$}\right)\right|\quad\textrm{for }z\in(0,1)\,.
\ee
Let us proceed to solve this problem. The first thing to notice is that for a fixed $f(z)$, the ratio is minimized by the $\tilde{g}(z)$ saturating the inequality. Therefore
\be
\tilde{g}(z) = (1-z)^{-2}\left|f\left(\mbox{$\frac{1}{1-z}$}\right)\right|\quad\textrm{for }z\in(0,1)
\label{eq:gTildeOptimal}
\ee
in the optimal functional. We find from \eqref{eq:actionLargeR}
\be
\omega_{\Df}(\Delta)\sim 2\mu(\Df,s)\sin^2\!\left[\frac{\pi}{2}\!\left(\Delta-2\Df + \delta\!\left(\mbox{$\frac{s}{4}$}\right)\right)\right]\left|f\left(\mbox{$\frac{s}{4}$}\right)\right|\,.
\label{eq:actionLargeR2}
\ee
We see that the functional has developed double zeros at
\be
\Delta = 2\Df + 2n - \delta\!\left(\mbox{$\frac{s}{4}$}\right)\,,
\label{eq:spectrumLargeR}
\ee
where $-\pi\delta(z)$ is the complex phase of $f(z)$. It remains to find the optimal $f(z)$ by minimizing the ratio
\be
 \frac{\int_{1}^{\infty}\!dz\,|f(z)|}{\mathrm{Im}[f(z_b)]}\,,
 \label{eq:optProblem}
\ee
subject to the usual constraints on the $f$ weight function, including the gluing condition. This optimization is an interesting problem in complex analysis which we solve in detail in appendix \ref{app:fOptimize}. The result is
\be
\boxed{
f(z) = \frac{2z-1}{[z(z-1)]^{1/2}(z-z_b)(z-1+z_b)S(z)}}\,,
\label{eq:fOptimal}
\ee
where
\be
S(z) =\pm\frac{\sqrt{z(1-z)}+\sqrt{z_b(1-z_b)}}{\sqrt{z(1-z)}-\sqrt{z_b(1-z_b)}}\,.
\label{eq:fPhase}
\ee
The upper (lower) sign applies when $z_b \in (0,1)$ is larger (lower) than 1/2, respectively. Note that when $z\in(1,\infty)$, $S(z)$ has unit modulus, and since all the other factors in $f(z)$ are real and positive, we can write
\be
S(z+i0^+) = e^{i\pi\delta(z)}\quad\textrm{for }z\in(1,\infty)\,,
\ee
where $\delta(z)$ controls the location of the double zeros through \eqref{eq:spectrumLargeR}.

Finally, we can also evaluate the ratio $-\omega_{\Df}(0)/\omega_{\Df}(m\Df)$ on the optimal solution, giving us the optimal upper bound on $c_{b}^2$. We find
\be
\boxed{
c_{b}^2\leq
\sqrt{64\pi\Df}
\frac{m^{3/2}\sqrt{2-m}}{\left|m^2-2\right| \sqrt{2+m}}
\left[\frac{2^{2(m+2)}}{
\left(2-m\right)^{2-m}\left(2+m\right)^{2+m}}\right]^{-\Df}
}\,,
\label{eq:cBoundSG}
\ee
where the bound is valid asymptotically as $\Df\rightarrow\infty$. 
We can see that for $m\neq\sqrt{2}$, the OPE coefficient must be exponentially supressed as $\Df\rightarrow\infty$. Our bound is singular for $m=\sqrt{2}$, signalling that the ansatz \eqref{eq:fglargeR} must be modified in this case, and the asymptotic behaviour of the bound will be different.

\subsection{Holographic scattering interpretation}
Having found the extremal functional, let us move on to identifying the corresponding extremal solution to crossing. The asymptotic bound that we found \eqref{eq:cBoundSG} exactly agrees with the results of \cite{Paulos:2016fap}. There the same conformal bootstrap problem was studied numerically at an increasing sequence of values of $\Df$ and for varying $m\in(0,2)$. The result was that the asymptotic behaviour of the upper bound as $\Df\gg1$ is well approximated by \eqref{eq:cBoundSG}.

In that reference it was shown that the corresponding asymptotic solution to crossing has a nice physical interpretation. It was identified as the boundary four-point function of the elementary $\Phi$ field in the sine-Gordon theory, placed in $AdS_2$ with large radius. To construct the theory, we can start with the free scalar of mass $M_\Phi$ placed in $AdS_2$ of radius $R$. Choosing the Dirichlet boundary condition on the bulk $\Phi$ field, the scaling dimension of the corresponding boundary operator $\phi(x)$ is given by the larger root of $\Df(\Df-1) = (M_\Phi R)^2$, and hence $\Df\sim M_\Phi R$ as $R\rightarrow\infty$. The $\phi\times\phi$ OPE in the free theory contains only the identity and the bilinear operators $\phi\!\!\overleftrightarrow{\partial}^{\!\!2n}\!\phi$ with scaling dimensions $\Delta_n = 2\Df + 2n$ where $n=0,1,\ldots$. In the $R\rightarrow\infty$ limit, these become the continuum of two-particle states of the theory in flat space.

We can now deform the bulk Lagrangian by a general unitary and UV-complete interaction. The interaction couplings can depend on $R$, but we would like to require that they approach the couplings of the sine-Gordon theory as $R\rightarrow\infty$:
\be
V(\Phi)\stackrel{R\rightarrow\infty}{\sim} M_{\Phi}^2\sum\limits_{n=1}^{\infty}(-1)^{n+1}\frac{g^{2n-2}}{(2n)!}\Phi^{2n}\,.
\label{eq:sgLagrangian}
\ee
The excitation sourced by the $\Phi$ field is the lightest breather of the sine-Gordon theory. The flat-space two-to-two exact S-matrix of the lightest breathers was computed long ago in \cite{Arefeva:1974bk,Zamolodchikov:1977py} with the result
\be
\mathcal{S}(s) =
\frac{\sqrt{s(4-s)}+m\sqrt{4-m^2}}{\sqrt{s(4-s)}-m\sqrt{4-m^2}}\,,
\label{eq:sgSmatrix}
\ee
where $s=(p_1+p_2)^2/M_{\Phi}^2$ and $m$ is the mass of the second-lightest breather in the units of $M_\Phi$.\footnote{$m$ depends on the coupling $g$ appearing in the Lagrangian as $m = 2\cos\!\left(\frac{\pi g^2}{16\pi - 2g^2}\right)$.} The second-lightest breather exists for $0<g^2<8\pi/3$, corresponding to the range $\sqrt{2}<m<2$. In this regime, the pole of the S-matrix at $s=m^2$ means the second-lightest breather can be thought of as a bound state of a pair of the lightest breathers. The only intermediate states in the $\Phi\Phi\rightarrow\Phi\Phi$ scattering process are this bound state and a continuum of two-particle states with $s>4$. Integrability of the theory means no three-, four- or higher-particle states can appear as intermediate states.

Let us translate these results to $AdS$. The intermediate states correspond to operators in the $\phi\times\phi$ OPE. The existence of the flat-space bound state of mass $mM_{\Phi}$ means that as long the bulk Lagrangian behaves as in \eqref{eq:sgLagrangian}, the $\phi\times\phi$ OPE will contain a primary operator $\mathcal{O}_b$ whose dimension behaves as $\Delta_b\sim m\Df$ as $\Df\rightarrow\infty$. Its OPE coefficient squared at large $\Df$ is then given precisely by the right-hand side of \eqref{eq:cBoundSG}. It can be computed from the geodesic tree-level exchange diagram in $\textrm{AdS}_2$, where the vertices contribute a factor proportional to the residue of the S-matrix \eqref{eq:sgSmatrix} at $s = m^2$ and the geodesics contribute a factor $e^{-M\times L}$ with $L$ the length of the geodesic. The latter factor is reponsible for the exponential suppression of $c_b^2$ as $\Df\rightarrow\infty$.

Besides the bound state with $\Delta\sim m\Df$, the $\phi\times\phi$ OPE will also contain an infinite tower of two-particle states. It can be shown (see \cite{Paulos:2016fap}) that as $\Df\rightarrow\infty$, their scaling dimensions are shifted from the free scalar values $2\Df + 2n$ by the phase of the flat-space S-matrix. Indeed, let us write the sine-Gordon S-matrix in the two-particle regime $s>4$ as $\mathcal{S}(s) = e^{i\pi\delta(s/4)}$. The scaling dimensions of the two-particle states for which $\Delta\sim \sqrt{s}\Df$ as $\Df\rightarrow\infty$ are given by
\be
\Delta = 2\Df + 2n - \delta\!\left(\mbox{$\frac{s}{4}$}\right)\,.
\label{eq:deltaShifts}
\ee
This is the same as the sequence of asymptotic double zeros of our extremal functional \eqref{eq:spectrumLargeR}, since the phase of $f(z)$ is minus the phase of the sine-Gordon S-matrix. Indeed, we have
\be
\mathcal{S}(s) = S\!\left(\mbox{$\frac{s}{4}$}\right)
\ee
with $S(z)$ given in \eqref{eq:fPhase}.

It should be noted that at any finite radius $R$, i.e. finite $\Df$, the $\phi\times\phi$ OPE will generically contain other primary operators corresponding to composites of the form $[\phi^{2k}]$ with $k=2,3,\ldots$. However, provided the Lagrangian approaches the flat-space integrable theory \eqref{eq:sgLagrangian} as $R\rightarrow\infty$, the contribution of these operators to the four-point function will be subleading compared to that of the bound state and the two-particle states in the large-radius limit.
 
This finishes the identification of the asymptotic extremal solution to crossing in the regime $\sqrt{2}<m<2$, thus showing that we have indeed constructed the optimal asymptotic functional. It would be interesting to identify the extremal theory (or rule out its existence) in the regime $0<m<\sqrt{2}$.

\subsection{S-matrix bootstrap from the conformal bootstrap}\label{ssec:smBootstrap}
The problem we analyzed has a direct analogue in the framework of the flat-space S-matrix bootstrap, considered in \cite{Creutz:1973rw,Paulos:2016but}. Since flat space can be thought of as the $R\rightarrow\infty$ limit of $AdS$, we expect that any S-matrix bootstrap result should at least in principle be derivable from an appropriate limit of the conformal bootstrap equations in one fewer spacetime dimensions. The purpose of this section is to illustrate how one can derive some results of the 2D S-matrix bootstrap from our approach to the 1D CFT bootstrap at large $\Df$.

Before reviewing the S-matrix bootstrap problem and its solution, let us generalize the OPE maximization problem to include multiple bound states. We allow for the $\phi\times\phi$ OPE to include primary operators with dimensions $\Delta_j$, $j=1,\ldots,N$ such that $\Delta_j\sim m_j\Df$ as $\Df\rightarrow\infty$, where $m_j\in(0,2)$ are kept fixed. We want to maximize the OPE coefficient of a fixed bound state labeled $k$.

We would like to determine the optimal $f(z)$ and $\tilde{g}(z)$ corresponding to the new problem. The only difference compared to the analysis of section \ref{ssec:solution} is that we now also need to impose $\omega_{\Df}(m_j\Df) \geq 0$ for all bound states. Thanks to \eqref{eq:fActionOnBound}, this is equivalent to imposing $\mathrm{Im}[f(z_j)]>0$ for all $j$, where $z_j = m_j^2/4$. For a given $f(z)$, the bound is still optimized by $\tilde{g}(z)$ satisfying \eqref{eq:gTildeOptimal} and it remains to determine the optimal $f(z)$.

Following the discussion in appendix \ref{app:fOptimize}, we can eliminate the branch cut of $f(z)$ in $z\in(0,1)$ by writing
\be
f(z) = -\frac{2z-1}{[z(z-1)]^{1/2}}f_{1}(z)\,,
\ee
where $f_1(z) = f_1(1-z)$. To satisfy the positivity criterium on the bound states, $f_1(z_j)$ should have the same sign as $2z_j-1$ for all $j$. Let us make the following change of variables mapping the cut plane to the unit disk
\be
x(z) = \frac{\sqrt{z_k(1-z_k)}-\sqrt{z(1-z)}}{\sqrt{z_k(1-z_k)}+\sqrt{z(1-z)}}\,,
\ee
so that $z_k$ gets mapped to $x=0$ and the remaining bound state locations $z_j$ get mapped to $x_j = x(z_j)\in(-1,1)$. Following appendix \ref{app:fOptimize}, let us write
\be
f_3(x) = \frac{f_1(z(x))}{1+x^2}\,,
\ee
so that the problem is equivalent to minimizing
\be
\frac{\frac{1}{2\pi}\!\int_{0}^{2\pi}\!\! d\theta\,|f_{3}\!\left(e^{i\theta}\right)\!|}{|f_3(0)|}
\label{eq:ratioToMax}
\ee
subject to $f_3(x_j)$ having the same sign as $2z_j - 1$ for all $j$. Let us choose the labels $j$ so that $x_j$ with $j=1,\ldots,N$ form an increasing sequence. If all $2z_j - 1$ were of the same sign, the optimization problem is solved by $f_3(x) = \pm 1$, as explained in appendix \ref{app:fOptimize}. In this case, the extremal functional has no additional zeros corresponding to exchanged operators in the region $0<\Delta<2\Df$. In other words, the optimal solution to crossing does not contain the bound states with $j\neq k$ and the bound on the OPE coefficient is unchanged compared to the case of a single bound state.

For a general configuration of $x_j$, the only complication arises when there are two consecutive bound states at $x_j$ and $x_{j+1}$ such that $2z_j-1$ and $2z_{j+1}-1$ have opposite signs. In that case, we need to insert a zero at a location $y_j\in(x_j,x_{j+1})$. Since we would like to introduce this zero without modifying the modulus of $f_3(x)$ on the unit circle, let us define the following function
\be
\alpha(x,y) =\frac{x-y}{x y - 1}\,.
\label{eq:alphaDef}
\ee
Importantly, $|\alpha_j(x,y)| = 1$ for $y\in\mathbb{R}$ and $x$ on the unit circle. For a given choice of the precise location of zeros $y_i$, the optimal $f_3(x)$ reads
\be
f_3(x) =\pm\prod\limits_{j\in D}\alpha(x,y_j)\,,
\ee
where $j$ runs over all the bound states such that $2z_j-1$ and $2z_{j+1}-1$ have opposite signs and $x_j<y_j<x_{j+1}$. We still have the freedom to optimize the $y_j$'s in order to minimize the ratio \eqref{eq:ratioToMax}. Since $\alpha(0,y) = y$, the minimal value is achieved when each $|y_j|$ in the product is maximized. This in turn happens when $y_j=x_j$ or $y_j=x_{j+1}$, whichever of the two has a greater absolute value. Hence, the final result is
\be
f_3(x) =\pm\prod\limits_{j\in \tilde{D}}\alpha(x,x_j)\,,
\label{eq:tildeDDef}
\ee
where the set $\tilde{D}$ consists of all $j$ such that the sign of $2z_j-1$ is the opposite from the sign of $2z_{j-\mathrm{sgn}(x_j)}-1$. The overall sign is chosen to ensure $f_3(0)$ has the same sign as $2z_k-1$.

We can see that the functional has developed zeros asymptotically at $m_j\Df$ with $j\in \tilde{D}$. Therefore, the asymptotic optimal solution to crossing will only contain bound states at these locations, together with the bound state at $m_k\Df$. Translating our result for the optimal $f_3(x)$ back to $f(z)$, we find
\be
\boxed{
f(z) = \frac{2z-1}{[z(z-1)]^{1/2}(z-z_k)(z-1+z_k)S(z)}}\,,
\ee
where
\be
S(z(x)) = \pm\frac{1}{x}\prod\limits_{j\in \tilde{D}}\alpha^{-1}(x,x_j)\,,
\label{eq:sMxOpt}
\ee
where the overall sign is chosen so that the residue at $x=0$ has the opposite sign than $2z_k-1$. $S(z)$ has poles at $z=m_j^2/4$ for each of the bound states appearing in the optimal solution to crossing. Just like in the case of a single bound state, $S(z)$ has unit modulus on the branch $z\in(1,\infty)$, and its phase there coincides with minus the phase of $f(z)$. Therefore, it has a natural interpetation as the S-matrix of a flat-space theory with no particle production \cite{Dorey:1996gd,Gabai:2018tmm}, with the two-particle scaling dimensions determined by the phase of $S(z)$ as in \eqref{eq:deltaShifts}.

Finally, the upper bound on the OPE coefficient $c_b^2$, given by $-\omega(0)/\omega(m_k\Df)$ is modified. $\omega(0)$ is the same as in the case of a single bound state since $S(z)$ has a unit modulus on the branch cut. On the other hand, $|\omega(m_k\Df)|^{-1}$ gets multiplied by the relative new factor in $\mathrm{Im}[f(z_k)]^{-1}$, i.e. by
\be
\prod\limits_{j\in\tilde{D}}\frac{1}{|x_j|}\,.
\ee
Since $|x_j|\in(0,1)$ for all $j$, the bound goes up. This could be expected since by introducing the possiblity for additional bound states, we enlarged the space of allowed solution to crossing.

This completes our discussion of asymptotic OPE maximization with multiple bound states-like primary operators. Let us proceed by reviewing the analogous set-up directly in the context of the 2D S-matrix bootstrap \cite{Paulos:2016but}. We will see that the final results will be equivalent but the way this happens is relatively nontrivial.

One considers a 2D S-matrix $\mathcal{S}(s)$ of the two-to-two scattering of identical particles, assumed to be the lightest particle of the theory. All dimensionful quantities are measured in the units of the scattered particle's mass. $\mathcal{S}(s)$ is meromorphic in $\mathbb{C}\backslash((-\infty,0)\cup(4,\infty))$ and satisfies crossing symmetry and unitarity
\be
\mathcal{S}(s) = \mathcal{S}(4-s)\quad\textrm{and}\quad |\mathcal{S}(s)|\leq 1\quad\textrm{for }s\in(4,\infty)\,.
\ee
We fix the spectrum of bound state masses to be $m_j$, $j=1,\ldots,N$, where $0<m_j<2$. This means $\mathcal{S}(s)$ has simple poles with negative residues at $s=m_j^2$ and simple poles with positive residues at $s=4-m_j^2$. $\mathcal{S}(s)$ is holomorphic away from these poles in the cut plane. The question is to identify the S-matrix that maximizes the coupling of the scattered particle to the $k$th bound state, i.e. with maximally negative residue at $s=m_k^2$.

This problem admits an analytic solution, derived in \cite{Creutz:1973rw} for the case of a single bound state and in \cite{Paulos:2016but} with multiple bound states, which we will now review. The first step is to make a familiar change of variables
\be
x(s) = \frac{m_k\sqrt{4-m_k^2}-\sqrt{s(4-s)}}{m_k\sqrt{4-m_k^2}+\sqrt{s(4-s)}}\,,
\ee
The cut plane gets mapped to the interior of the unit disk and the point $s=m_k^2$ to $x=0$. A pole at $s=m_j^2$ with negative residue gets mapped to a pole at $x_j = x(m_j^2)\in(-1,1)$ with residue of the same sign as $2-m_j^2$. Let us denote $\tilde{\mathcal{S}}(x) = \mathcal{S}(s(x))$. We will start with the case where there is a single bound state, with mass $m_k$, whose coupling we want to maximize. $\tilde{S}(x)$ is bounded by $1$ on the unit circle and its only singularity is a simple pole at $x=0$. Hence $x\tilde{\mathcal{S}}(x)$ is holomorphic inside the unit circle, bounded by 1 on the unit circle and we want to maximize its absolute value at $x=0$. Hence the solution is
\be
\tilde{\mathcal{S}}(x) = \pm\frac{1}{x}\,,
\ee
where the sign is the same as the sign of $2-m_k^2$. When $m_k>\sqrt{2}$, this is precisely the S-matrix of the lightest breathers of the sine-Gordon theory.

In order to solve the problem with a general spectrum of bound states, let us again consider $\tilde{\mathcal{S}}(x)$. Let us choose the labels $j$ so that $x_{j}$ with $j=1,\ldots,N$ form an increasing sequence. For each $j$, $\tilde{\mathcal{S}}(x)$ should have a pole at $x_j$ whose residue has the same sign as $2-m_j^2$.

We would like to introduce the additional poles of $\tilde{\mathcal{S}}(x)$ without changing the modulus on the unit circle. To achieve that, we can again use $\alpha(x,x_j)$ from \eqref{eq:alphaDef} and write the following ansatz
\be
\tilde{\mathcal{S}}(x) =A(x) \prod\limits_{j=1}^{N}\alpha^{-1}(x,x_j)\,,
\ee
so that $A(x)$ is holomorphic inside the unit disk and bounded by one on the unit circle. However, $A(x)$ can not be a constant in general since the product accompanying it has residues that alternate in sign. If the sequence of $m_j^2-2$ with $j=1,\ldots,N$ has alternating signs, there is no issue and we can set $A(x) = \pm 1$ to get the optimal S-matrix, where the sign depends on the distribution of $m_j^2$ around 2. On the other hand, when there is $j$ such that $m_j^2-2$ and $m_{j+1}^2-2$ have the same sign, we need to make sure $\tilde{\mathcal{S}}(x)$ has a zero in between $x_j$ and $x_{j+1}$. To do that, we simply insert $\alpha(x,y_j)$ with $x_j<y_j<x_{j+1}$ into the product. Doing this for all consecutive pairs of bound states with equal sign of $m_j^2-2$, we end up with the ansatz
\be
\tilde{\mathcal{S}}(x) =B(x) \prod\limits_{j=1}^{N}\alpha^{-1}(x,x_j)\prod\limits_{i\in C}\alpha(x,y_i)\,,
\ee
where $C$ includes all $j$ such that $m_j^2-2$ and $m_{j+1}^2-2$ have the same sign. Given a fixed choice of all $y_i$s, the residue at $x=0$ is now maximizied by $B(x) =\pm 1$. Moreover, the values $y_i$ can be varied to maximize the residue at $x=0$ too. This residue is proportional to $\prod_{i\in C}y_i$. Therefore, it is maximized if all $y_i$ are taken as far from $x=0$ as possible, i.e. precisely cancelling the pole at $x_{j+\mathrm{sgn}(x_j)}$. We conclude the optimal S-matrix takes the form
\be
\tilde{\mathcal{S}}(x) =\pm\frac{1}{x}\prod\limits_{j\in \tilde{D}}\alpha^{-1}(x,x_j)\,,
\label{eq:sMatrixOpt}
\ee
where the set $\tilde{D}$ is exactly the same as the one needed to optimize the bootstrap functional, see \eqref{eq:tildeDDef}. We can see that the optimal S-matrix coincides with the S-matrix arising from the conformal bootstrap problem \eqref{eq:sMxOpt} after the substitution $z=s/4$. Moreover, the maximal coupling, i.e. the residue of the optimal S-matrix at $s=m_k^2$ is obtained from the maximal coupling with no extra bound states by multiplying by the factor
\be
\prod\limits_{j\in D}\frac{1}{|x_j|}\,,
\label{eq:newFactorBound}
\ee
again agreeing with the conformal bootstrap prediction. This completes our demonstration that in this context, the S-matrix bootstrap results can be derived from the conformal bootstrap in the large-$\Delta$ limit.

We conclude this section by making a suggestive observation. At first, it may seem peculiar that the functional contains the S-matrix in the denominator. However, there is a different way to think about it. Consider again the map to the unit disk $z\to x(z)$. Since along the boundary of the disk we have $|\tilde{\mathcal S}(x)|=1$ we can define the analytic continuation outside the disk by $\tilde {\mathcal S}(1/x)=1/\tilde{\mathcal S}(x)$. Back in the $z$ variable the disk exterior corresponds to a second copy of the complex plane obtained by traversing the cuts. So really we should think of the functional as naturally living on the second sheet of the Mandelstam plane, where  the poles of the S-matrix become zeros. In this way the functional and the S-matrix are unified between the sheets.

\section{Conclusions and outlook}
In this work, we have studied a class of linear functionals that act on the conformal bootstrap equation arising from the crossing symmetry of the four-point function of identical operators in 1D CFTs. We have argued that these functionals are ideally suited for extracting information from the crossing equation, in the form of bounds on CFT data, thereby extending the philosophy first set out in \cite{Rattazzi:2008pe} from the realm of numerics to analytics. The functionals take the form of integrals in the complexified cross-ratio space against suitable weight functions. The weight functions satisfy certain analyticity properties and non-trivial functional equations.

While solving the optimization problem for the weight function analytically in general remains a formidable task, we have shown that in certain simplified settings exact solutions can be found. In particular, we have shown our ansatz is sufficiently general to capture the extremal functionals for the gap maximization in 1D, as well as a class of OPE maximization functionals, where the optimal bounds are saturated by the generalized free fermion, extending the results in \cite{Mazac:2016qev}. More interestingly, we have solved the optimization problem exactly for OPE maximization at large conformal dimension. We found that the optimal solutions to crossing then correspond to holographic duals of 2D integrable field theories placed in large $AdS_2$, thus analytically establishing the results of \cite{Paulos:2016fap}.

Since we solved the OPE maximization problem exactly at the leading order at large $\Df$, it would be very interesting to use this as a starting point of a perturbative analysis around infinite $\Df$. In the context of field theories in $AdS$, the $1/\Df$ corrections come from the $AdS$ space having a finite radius $R$.  The solution of the OPE maximization problem at large but finite $\Df$ presumably corresponds to a distinguished theory of a single scalar field in $AdS_2$, whose couplings approach those of the flat-space sine-Gordon model as $\Df\rightarrow\infty$. The $\phi\times\phi$ OPE in a generic field theory in large but finite $AdS_2$ will contain primary states corresponding to four- six- and higher $\phi$-particle states. When the couplings approach those of a flat-space integrable theory as $R\rightarrow\infty$, the contribution of these states to the asymptotic four-point function decouples in this limit and we are left with only the bound state and two-particle states. 

However, the numerical bootstrap indicates that the solution to crossing maximizing the OPE coefficient of the bound state only contains the bound state and two-particle states even at \emph{finite} $\Df$, see \cite{Paulos:2016fap}. This requires a large amount of fine-tuning of the bulk couplings as a function of $R$. Note that such a theory could not exist in more than two (boundary) dimensions since multi-twist composites of $\phi$ must always be present in the OPE. It should be possible to identify the theory perturbatively in $1/R$ both using our method and using direct computation starting from a general bulk Lagrangian. We would thus obtain an interesting two-parameter family of solutions to crossing, parametrized by $\Df$ and $\Delta_b$ from the point of view of the 1D CFT and by $R$ and the sine-Gordon coupling $g$ from the point of view of the bulk. Drawing a rough analogy with the more complicated case of string theory in the bulk of $AdS$, our operator $\phi$ would correspond to a massive string state, and the parameters $R$ and $g$ correspond to (a power of) $\lambda$ and $1/N$ respectively. 

The functionals we have constructed are extremal, i.e. they automatically come with associated exact solutions to crossing. In such cases, it is known that at least in the truncated, numerical context, one can use the functionals to construct flows in the space of CFT data starting at the original solution that remain crossing-symmetric along the flow \cite{El-Showk2016}. One can use our functionals to find these flows analytically, but this will require an infinite set of functionals telling us how individual OPE coefficients and scaling dimensions vary along the flow. These and other matters will be explored in an upcoming work \cite{Mazac:2018ycv}.

One should attempt to generalize our method to other contexts where numerical bootstrap has proven powerful, such as in higher dimensions, in the presence of global symmetries and with mixed correlators. The boundary bootstrap of \cite{Liendo2013} and modular bootstrap can plausibly also be tackled with our approach.

Finally, it would be interesting to understand whether the SYK or related models can saturate appropriate bootstrap bounds. The tools of this work should prove relevant in that context.

\subsection*{Acknowledgements}
It is a pleasure to acknowledge useful discussions with L. C\'{o}rdova, L. Di Pietro, D. Gaiotto, M. Hogervorst, Z. Komargodski, S. Komatsu, L. Rastelli, S. Rychkov, B. van Rees, and P. Vieira. We thank the organizers and participants of the Simons Non-perturbative Bootstrap Workshop and School at the ICTP-SAIFR in Sao Paulo, where part of this work was completed, for creating a stimulating research environment. DM is grateful to the organizers and participants of the OIST Symposium: ``Bootstrap Approach to Conformal Field Theories and Applications'' for providing an inspiring environment during the completion of this work.

\appendix
\section{Details on the free functionals}
This appendix contains technical details omitted from the main text, which concern the functionals for the generalized free fermion.
\subsection{A lower bound}\label{app:boundN}
Firstly, let us prove that the integer $n$ in \eqref{eq:z1behaviour} is bounded by $n\geq 0$ under our assumptions on $f(z)$. Let us recall that $f(z)$ is analytic in $\mathbb{C}\backslash[0,1]$, satisfies $f(z) = f(1-z)$, $f(\bar{z}) = \overline{f(z)}$ and is holomorphic at $z=\infty$ where it must decay at least as $z^{-2}$. There are no singularities in $z\in(0,1)$. Furthermore, $f(z)<0$ for $z\in(1,\infty)$. Finally $f(z)$ is constrained by the fundamental relation \eqref{eq:fundamentalfree}.

Let us map the region $z\in\mathbb{C}\backslash[0,1]$ to the interior of the unit disk $|w|<1$ by
\be
z(w) = \frac{(1+w)^2}{4w}\,.
\ee
The branch cut $z\in[0,1]$ gets mapped to the unit circle, and $z=\infty$ to $w=0$. Define
\be
\tilde{f}(w) = f(z(w))\,.
\ee
$\tilde{f}(w)$ is holomorphic in the unit disk with possible exceptions at $w=\pm 1$ and satisfies $\tilde{f}(w) = \tilde{f}(-w)$. Furthermore, we have
\be
\tilde{f}(0) = \lim_{z\rightarrow\infty}f(z) = 0\,.
\ee
Suppose, seeking contradiction, that $n < 0$. In this case the singularity at $w=\pm1$ is at most logarithmic, and we can use holomorphy of $\tilde{f}(w)$ to write its value at the origin as the average over the unit circle
\be
0 = \tilde{f}(0) = \frac{1}{2\pi}\int\limits_{0}^{2\pi}\!\!d\theta\tilde{f}(e^{i\theta})\,.
\ee
This integral can be further simplified by using the symmetry and reality of $\tilde{f}(w)$
\be
0 = \frac{2}{\pi}\!\int\limits_{0}^{\pi/2}\!\!d\theta\,\mathrm{Re}[\tilde{f}(e^{i\theta})]\,.
\label{eq:contradiction}
\ee
The fundamental relation \eqref{eq:fundamentalfree} relates the real part of $f(z)$ on the branch cut to its values for $z\in(1,\infty)$, where it is non-positive by our assumption. Therefore, \eqref{eq:contradiction} cannot be satisfied, showing that the singularity at $z = 1$ must be stronger, or $n\geq 0$.

\subsection{Mellin inversion}\label{app:mellin}
In Section \ref{sec:construct}, we defined the following Mellin-like transform of the weight-function $f(z)$
\be
M(s) = -\frac{1}{2\cos(\pi s)}\int\limits_{0}^1\!\!dz \,[z(1-z)]^s\mathrm{Re}[f(z)].
\label{eq:mellinDefApp}
\ee
Here we would like to show how to invert this transform. Recall that for the normal functional, $f(z)\sim -2\pi^{-2}(z-1)^{-2}$ for $z\to 1$. Using the fundamental relation \reef{eq:fundamentalfree} it follows that $\cos(\pi s)M(s)$ is holomorphic for $\mathrm{Re}(s)>1$ and has a simple pole at $s=1$. It will be convenient to define $\tilde{f}(z)$ with softer behaviour at $z=0,1$ as follows
\be
\tilde{f}(z) = f(z) +\frac{2}{\pi^2z^2(1-z)^2}\,,
\ee
so that $\tilde{f}(z)\!\stackrel{z\rightarrow 1}{=}\!o((z-1)^{-1-\epsilon})$ for any $\epsilon > 0$. Let us also define the analogous transform of $\tilde{f}(z)$
\be
\tilde{M}(s) = -\frac{1}{2\cos(\pi s)}\int\limits_{0}^1\!\!dz \,[z(1-z)]^s\mathrm{Re}[\tilde{f}(z)]\,.
\label{eq:mTilde}
\ee
We have
\be
\tilde{M}(s) = M(s) - \frac{\Gamma(s-1)^2}{\pi^2\cos(\pi s)\Gamma(2s-2)}\,.
\ee
The upshot is that the integral in \eqref{eq:mTilde} converges for $\mathrm{Re}(s)>0$ and not just for $\mathrm{Re}(s)>1$. Now, provided that $0<\mathrm{Re}(s)<1/2$, we can deform the integration contour in \eqref{eq:mTilde} and obtain
\be
\tilde{M}(s) = \int\limits_{1}^{\infty}\!\!dz \,[z(z-1)]^s\tilde{f}(z)\,.
\ee
This integral becomes a standard Mellin transform after the change of variables $w=z(z-1)$. It can be inverted as follows
\be
\tilde{f}(z) = \frac{2z-1}{z(z-1)}\!\!\int\limits_{s_0-i\infty}^{s_0+i\infty}\!\!\frac{ds}{2\pi i} \,[z(z-1)]^{-s}\tilde{M}(s)\,,
\label{eq:fTildeInv}
\ee
where $0<s_0<1/2$. Now, note that
\ba
f(z) - \tilde{f}(z) &= 
-\frac{2}{\pi^2z^2(z-1)^2} =\\
& =\frac{2z-1}{z(z-1)}\!\int\limits_{\Gamma}\!\frac{ds}{2\pi i} \,[z(z-1)]^{-s}
\left[\frac{\Gamma(s-1)^2}{\pi^2\cos(\pi s)\Gamma(2s-2)}\right] = \\
& = \frac{2z-1}{z(z-1)}\!\int\limits_{\Gamma}\!\frac{ds}{2\pi i} \,[z(z-1)]^{-s}
\left[M(s) - \tilde{M}(s)\right]
\,,
\ea
where $\Gamma$ passes to the left of the poles at $s=1/2 + n$, $n=0,1,\ldots$ but to the right of the pole at $s=1$. Since the contour in \eqref{eq:fTildeInv} can be deformed to become $\Gamma$, we finally arrive at the inversion formula
\be
f(z) = \frac{2z-1}{z(z-1)}\!\int\limits_{\Gamma}\!\frac{ds}{2\pi i} \,[z(z-1)]^{-s}M(s)\,.
\ee

\subsection{A differential equation for $f(z)$}\label{app:difeq}
Consider the following highly symmetric third-order linear homogenous ODE for an unknown function $f(z)$
\ba
f^{(3)}(z)&\phantom{[}2 (z-2) (z-1)^2 (z+1) (2 z-1) z^2+\phantom{]}\\
+ f''(z)&\left[4 (z-1) z \left(7 z^4-14 z^3-9 z^2+16 z-5\right)-\right.\\
&-\left.4 (z-2) (z-1) z (z+1) (2 z-1)^2 \Delta _{\phi }\right]+\\
+f'(z)&\left[6 (2 z-1) \left(3 z^4-6 z^3-8 z^2+11 z-2\right)-\right.\\
&-\left.4 (2 z-1) \left(8 z^4-16 z^3-11 z^2+19 z-6\right) \Delta _{\phi }+\right.\\
&+\left.8 (z-2) (z-1) z (z+1) (2 z-1) \Delta _{\phi }^2\right]+\\
+f(z)&\left[4 (z-2) (z+1) \left(8 z^2-8 z+3\right) \Delta _{\phi }^2-3 \left(z^2-z+4\right)-\right.\\
&-\left.4 \left(8 z^4-16 z^3-32 z^2+40 z-9\right) \Delta _{\phi }\right]= 0\,.
\ea
The equation has regular singular points at $z=0,1,\infty$ and $z=-1,1/2,2$. The ODE is invariant under the group $S_3$ permuting the above triples of points generated by the following transformations
\ba
f(z)&\mapsto f(1-z)\\
f(z)&\mapsto z^{2\Df-2}f(1/z)\,.
\ea
Since $z=\infty$ is a regular singular point, the solution space is generated by solutions with leading behaviour $z^\alpha$ as $z\rightarrow\infty$. The allowed values are $\alpha = -2,2\Df-2,2\Df$. Taking $\Df>0$, there is a unique solution $f(z)$ (up to an overall constant) such that $f(z)=O(z^{-2})$ for large $z$. This asymptotic behaviour is invariant under the symmetry $z\mapsto1-z$, and hence the solution must satisfy $f(z)=f(1-z)$. It can be checked, for example by a series expansion around $z=\infty$, that the proposal for $f(z)$ given in \eqref{eq:allaf} solves the ODE for general $\Df>0$ and therefore must agree with this unique solution. Our goal is to show using the ODE that the fundamental relation \eqref{eq:fundamentalfree} holds for $f(z)$.

We start by considering $g(z)$ defined from $f(z)$ as in \eqref{eq:gFree}
\be
g(z) = - (1-z)^{2\Df - 2}f\!\left(\mbox{$\frac{1}{1-z}$}\right)\,.
\ee
Thanks to a symmetry of the ODE, $g(z)$ is also its solution. Furthermore, thanks to the symmetry of the ODE under $z\leftrightarrow 1-z$, $g(z) + g(1-z)$ is yet another solution. This is the unique solution (up to a constant) of the ODE which is regular at $z=1/2$ and symmetric under $z\leftrightarrow 1-z$. The way to see this is as follows. $z=1/2$ is a regular singular point where the leading power-law behaviour of a solution must be $(z-1/2)^\alpha$ with $\alpha = 0,1,3$. The only symmetric solution thus has $\alpha = 0$ and so $g(z) + g(1-z)$ must be this solution. We would like to show that
\be
g(z)+g(1-z) = -\mathrm{Re}[f(z)]\quad\textrm{for }z\in(0,1)\,.
\ee
$f(z)$ has a branch cut for $z\in(0,1)$ so consider instead $\tilde{f}(z)$ defined as $f(z)$ for $\mathrm{Im}(z)>0$ and by an analytic continuation through the interval $(0,1)$ to $\mathrm{Im}(z)<0$. Symmetry of $f(z)$ under $z\mapsto1-z$ implies that
\be
2\,\mathrm{Re}[f(z)] = \tilde{f}(z) + \tilde{f}(1-z)\quad\textrm{for }z\in(0,1)\,.
\ee
Note that $\tilde{f}(z)$ has branch cuts for $z\in(-\infty,0)\cup(1,\infty)$, just like $g(z) + g(1-z)$. Now $\tilde{f}(z)+\tilde{f}(1-z)$ is symmetric under $z\mapsto1-z$ and regular at $z=1/2$, but we know $g(z)+g(1-z)$ is the unique such solution up to a scalar multiplication. Therefore the two must be proportional to each other
\be
\tilde{f}(z)+\tilde{f}(1-z) = a \left[g(z) + g(1-z)\right]
\label{eq:fTildeSym}
\ee
for some $a\in\mathbb{C}$. The fundamental relation follows if we can show $a=-2$. $a$ can be fixed by looking at the coefficient of the double pole at $z=1$ at both sides of the equation. Suppose
\be
f(z)\sim \frac{b}{(z-1)^2}\quad\textrm{as }z\rightarrow 1\,.
\ee
We know this is true from the explicit expression for $f(z)$ given in \reef{eq:allaf}. It is also, as we will see now a self-consistent assumption.
By symmetry $f(z)$ also has a double pole at $z=0$.
The analytic continuation $\tilde f(z)$ must have the same double poles with the same residue, and hence the coefficient of the double pole on the LHS of \eqref{eq:fTildeSym} is $2b$. Moreover
\ba
&g(z) = - (1-z)^{2\Df - 2}f\!\left(\mbox{$\frac{1}{1-z}$}\right) \stackrel{z\rightarrow1}{=} O((1-z)^{2\Df})\\
&g(1-z)= -  z^{2\Df - 2}f\!\left(\mbox{$\frac{1}{z}$}\right) \stackrel{z\rightarrow1}{\sim} -\frac{b}{(z-1)^2}\,.
\ea
Therefore, the constant in \eqref{eq:fTildeSym} is $a = -2$. We conclude $f(z)$ satisfies the fundamental relation.

\section{Details on the OPE maximization}
This appendix fills in some technical details in the discussion of section \ref{sec:largeR}.

\subsection{The functional for $\Delta > 2\Df$}\label{app:saddleHigh}
Here we evaluate the integral \eqref{eq:fAction5}
\be
\omega_{\Df}(\Delta) = \int\limits_{1}^{\infty}\!\!dz\!\left\{z^{2\Df - 2}\,g_{\Df}\!\left(\frac{z-1}{z}\right) - 
\mathrm{Re}\left[e^{-i\pi(\Delta-2\Df)}f_{\Df}(z)\right]\right\}\frac{\widehat{G}_{\Delta}(1-z)}{(z-1)^{2\Df}}
\ee
in the flat space limit, where $\Delta = \sqrt{s}\Df$ with $s>4$ and $\Df\rightarrow\infty$. We assume the asymptotic behaviour of $f_{\Df}(z)$ and $g_{\Df}(z)$ is as in \eqref{eq:fglargeR}, i.e.
\ba
f_{\Df}(z)&\sim f(z)\qquad\quad\qquad\textrm{for }\mathrm{Im}(z)>0\\
g_{\Df}(z)&\sim (1-z)^{2\Df}\tilde{g}(z)\,\,\;\textrm{for }z\in(0,1)\,.
\label{eq:fgAsympApp}
\ea
This implies that the factors inside the curly bracket approach the finite limit
\be
\left\{\phantom{\frac{A}{A}}\ldots\phantom{\frac{A}{A}}\right\}\sim z^{-2}\tilde{g}\!\left(\frac{z-1}{z}\right) - 
\mathrm{Re}\left[e^{-i\pi(\Delta-2\Df)}f(z)\right]\,.
\ee
On the other hand, the asymptotic behaviour of the factor outside the curly bracket reads
\be
\frac{\widehat{G}_{\sqrt{s}\Df}(1-z)}{(z-1)^{2\Df}} \sim
\frac{\sqrt{z}+1}{2z^{1/4}}
\left[2^{2\sqrt{s}}
\frac{ \left(\sqrt{z}-1\right)^{\sqrt{s}-2}}{\left(\sqrt{z}+1\right)^{\sqrt{s}+2}}\right]^{\Df}\,.
\ee
When $s>4$, as is the case for us, the expression in the square bracket has a unique stationary point in the region $z>1$: a maximum at $z=s/4$. Standard saddle-point approximation around this point then gives the result
\be
\omega_{\Df}(\Delta)\sim \mu(\Df,s)\left\{\left(\mbox{$\frac{s}{4}$}\right)^{-2}\tilde{g}\left(\mbox{$\frac{s-4}{s}$}\right)
-
\mathrm{Re}\!\left[e^{-i\pi(\Delta-2\Df)}f\left(\mbox{$\frac{s}{4}$}\right)\right]
\right\}\,,
\ee
where the prefactor reads
\be
\mu(\Df,s) =
\sqrt{\frac{\pi }{64\Df}}  s^{1/4}\!\left(\sqrt{s}+2\right)^{3/2}\!\left(\sqrt{s}-2\right)^{1/2}\!
\left[2^{2(\sqrt{s}+2)}\frac{
\left(\sqrt{s}-2\right)^{\sqrt{s}-2}}{\left(\sqrt{s}+2\right)^{\sqrt{s}+2}}\right]^{\Df}\,.
\ee

\subsection{The functional for $\Delta = \Delta_b$}\label{app:saddleLow}
Here we evaluate the action of the functional on the vector with $\Delta = \Delta_b\in(0,2\Df)$ in the large-radius limit where $\Delta_b = m \Df$ with $\Df \rightarrow\infty$. In this regime, the representation \eqref{eq:fAction} using an integral over $z\in(1,\infty)$ is not available since the integral does not converge and we need to go back to the original definition \eqref{eq:ffg} using a pair of contours
\be
\omega(\Delta) =\frac 12
\!\!\!\int\limits_{\frac{1}{2}}^{\frac{1}{2}+i\infty}\!\!\!\!dz\,f(z)\!
\!\left[\frac{G_{\Delta}(z)}{z^{2\Df}}-\frac{G_{\Delta}(1-z)}{(1-z)^{2\Df}}\right] + 
\!\!\int\limits_{\frac{1}{2}}^{1}\!\!dz\,g(z)\!
\!\left[\frac{G_{\Delta}(z)}{z^{2\Df}}-\frac{G_{\Delta}(1-z)}{(1-z)^{2\Df}}\right]\,.
\ee
Keeping $f(z)$ and $g(z)$ completely general, we can change variables in some factors to bring this to the form
\be
\omega(\Delta) =
-\left[\!\int\limits_{\frac{1}{2}}^{\frac{1}{2}+i\infty}\!\!\!\!dz\,\frac{f(z)}2
+\!\!\!\int\limits_{\frac{1}{2}}^{\frac{1}{2}-i\infty}\!\!\!\!dz\,\frac{f(1-z)}2
+\!\!\int\limits_{\frac{1}{2}}^{1}\!\!dz\,g(z)
+\!\!\int\limits_{\frac{1}{2}}^{0}\!\!dz\,g(1-z)\right]\frac{G_{\Delta}(1-z)}{(1-z)^{2\Df}}
\,,
\ee
where the factor outside the square bracket is meant to be distributed into each of the four integrands. Note that the gluing condition \eqref{eq:glue} can be stated as
\be
\frac{f(z) + f(1-z)}2 + g(z) + g(1-z) = 0\quad \textrm{for }z\in(0,1)\,.
\ee
where as usual the values of $f(z)$ are obtained by taking the limit from the upper-half plane.
This implies that we can simultaneously shift the lower limit of each of the integrals in the square bracket from $1/2$ to an arbitrary $z_0\in(0,1)$
\be
\omega(\Delta) =
-\left[\!\int\limits_{z_0}^{\frac{1}{2}+i\infty}\!\!\!\!dz\,\frac{f(z)}2
+\!\!\!\int\limits_{z_0}^{\frac{1}{2}-i\infty}\!\!\!\!dz\,\frac{f(1-z)}2
+\!\!\int\limits_{z_0}^{1}\!\!dz\,g(z)
+\!\!\int\limits_{z_0}^{0}\!\!dz\,g(1-z)\right]\frac{G_{\Delta}(1-z)}{(1-z)^{2\Df}}
\,.
\label{eq:intMoved}
\ee
When we set $\Delta = m\Df$ and take $\Df\rightarrow\infty$, the factor outside the square bracket behaves as follows
\be
\frac{G_{m\Df}(1-z)}{(1-z)^{2\Df}} \sim
\frac{\sqrt{z}+1}{2z^{1/4}}
\left[
\frac{2^{2m}}{\left(1-\sqrt{z}\right)^{2-m}\left(1+\sqrt{z}\right)^{2+m}}\right]^{\Df}\,.
\ee
When $0<m<2$ as is our case, the last expression has a saddle point at
\be
z_b = \frac{m^2}{4}\,.
\ee
The saddle point is a minimum when moving along the real axis, and the direction of the steepest descent is along the imaginary axis. Therefore, it is particularly convenient to set $z_0 = z_b$ in \eqref{eq:intMoved} so that each of the contours starts at the saddle point. We will now use the asymptotic behaviour of $f_{\Df}(z)$, $g_{\Df}(z)$ stated in \eqref{eq:fgAsympApp}. The contours in the first two terms of \eqref{eq:intMoved} can be chosen to run along the direction of the steepest descent. The exponential supression $(1-z)^{2\Df}$ of $g(z)$ implies that the last two terms are always subleading, leaving us with the saddle-point evaluation of the first two terms only
\be
\omega_{\Df}(m\Df)\!\sim\!
\frac{1}{16}\sqrt{\frac{\pi }{\Df}}  m^{1/2}\!\left(2+m\right)^{3/2}\!\left(2-m\right)^{1/2}\!
\left[\!\frac{2^{2(m+2)}}{
\left(2-m\right)^{2-m}\left(2+m\right)^{2+m}}\!\right]^{\Df}
\!\!\!\!\!\!\!\mathrm{Im}\!\left[f(z_b)\right]
\ee
where we used the symmetry and reality of $f(z)$ to write
\be
\frac{f(z_b) - f(1-z_b)}{2i} = \mathrm{Im}\!\left[f(z_b)\right]\,.
\ee

\subsection{Action on identity}\label{app:largeRidentity}
Here we will compute the action of the functional $\omega_{\Df}$ on identity in the limit $\Df\rightarrow\infty$. The representation \eqref{eq:fAction} does not directly apply for $\Delta = 0$. Instead, we can use a trick relying on the crossing symmetry of the generalized free fermion four-point function. The crossing symmetry implies that for any consistent functional $\omega$, we must have
\be
\omega(0) = - \sum\limits_{n=0}^{\infty}c_n^2\,\omega(2\Df + 2n + 1)\,,
\label{eq:idTrick}
\ee
where $c^2_n$ arise in the OPE decomposition of the generalized free fermion four-point function
\be
\sum\limits_{n=0}^{\infty}c_n^2\, G_{2\Df + 2n + 1}(z) = \left(\frac{z}{1-z}\right)^{2\Df}\!\!\! - z^{2\Df}\,.
\ee
Now, the representation \eqref{eq:fAction} applies to each term on the rhs of \eqref{eq:idTrick}. Therefore, we can write
\be
\omega_{\Df}(0) = -\int\limits_{1}^{\infty}\!\!dz\!\left\{z^{2\Df - 2}\,g_{\Df}\!\left(\frac{z-1}{z}\right) + 
\mathrm{Re}\!\left[f_{\Df}(z)\right]\right\}\sum\limits_{n=0}^{\infty}c_n^2\frac{\widehat{G}_{2\Df+2n+1}(1-z)}{(z-1)^{2\Df}}\,,
\ee
i.e. the sum over $n$ completely decouples from the part depending on $f_{\Df}(z)$. The infinite sum can be easily evaluated to give
\be
\omega_{\Df}(0) = -\int\limits_{1}^{\infty}\!\!dz\!\left\{z^{2\Df - 2}\,g_{\Df}\!\left(\frac{z-1}{z}\right) + 
\mathrm{Re}\!\left[f_{\Df}(z)\right]\right\}
\left(1 - z^{-2\Df}\right)\,.
\ee
As $\Df\rightarrow\infty$, both terms in the curly bracket approach a finite limit. On the other hand, the second term in the round bracket is subleading and we may drop it, finding
\be
\omega_{\Df}(0) \sim -\int\limits_{1}^{\infty}\!\!dz\!\left\{z^{-2}\,\tilde{g}\!\left(\frac{z-1}{z}\right) + 
\mathrm{Re}\!\left[f(z)\right]\right\}\,,
\ee
where we used the asymptotic properties of $f_{\Df}(z)$ and $g_{\Df}(z)$ stated in \eqref{eq:fgAsympApp}. Let us focus on the second term and rewrite it using the symmetry and reality of $f(z)$ as
\be
-2\!\!\int\limits_{1}^{\infty}\!\!dz\,
\mathrm{Re}\!\left[f(z)\right] = -\!\!\int\limits_{-\infty}^{0}\!\!dz\,f(z+i0^{+})
-  \!\!\int\limits_{1}^{\infty}\!\!dz\,f(z+i0^{+})\,.
\ee
Since $f(z)$ decays sufficiently fast as $|z|\rightarrow\infty$, we can add a semicircle at infinity at no cost and contract the contour to find
\be
-2\!\!\int\limits_{1}^{\infty}\!\!dz\,
\mathrm{Re}\!\left[f(z)\right] = \!\!\int\limits_{0}^{1}\!\!dz\,
\mathrm{Re}\!\left[f(z)\right] = 0\,.
\ee
where we used the $\Df\rightarrow\infty$ limit of the gluing condition to obtain the last equality. Therefore, we conclude
\be
\omega_{\Df}(0) \sim -\int\limits_{1}^{\infty}\!\!dz\, z^{-2}\,\tilde{g}\!\left(\frac{z-1}{z}\right) =
-\int\limits_{0}^{1}\!\!dz\,\tilde{g}(z)\,.
\ee
as $\Df\rightarrow\infty$.

\subsection{Optimizing $f(z)$}\label{app:fOptimize}
Our goal here will be to perform the final step of the optimization and thus determine the limiting function $f(z)$. Specifically, we would like to minimize
\be
 \frac{\int_{1}^{\infty}\!dz\,|f(z)|}{\mathrm{Im}[f(z_b)]}\,,
 \label{eq:toMaxApp}
\ee
where $z_b=m^2/4\in(0,1)$, and where $f(z)$ satisfies a number of constraints, which we summarize now. $f(z)$ is holomorphic away from the cut $z\in\mathbb{R}$ and on the cut is only allowed to have singularities at $z=0,1$ and $\infty$. $\mathrm{Im}[f(z_b)]$ is to be evaluated at $z_b+i0^+$. $f(z)$ satisfies the following symmetry and reality properties
\be
f(z) = f(1-z)\quad\textrm{and}\quad f(\bar{z}) = \overline{f(z)}\,.
\ee
$f(z)$ is not allowed to grow faster than $O((z-1)^{-1+\epsilon})$ for some $\epsilon>0$ as $z\rightarrow 1$, and must decay at least as $O(z^{-1-\epsilon})$ for some $\epsilon>0$ as $z\rightarrow \infty$. In particular, this guarantees the numerator in \eqref{eq:toMaxApp} is a convergent integral. Finally, we should remember $f_{\Df}(z)$ and $g_{\Df}(z)$ satisfy the gluing condition \eqref{eq:glue} for any finite $\Df$. Since $g_{\Df}(z)$ is exponentially supressed for $z\in(0,1)$ as $\Df\rightarrow\infty$, the gluing condition implies
\be
\mathrm{Re}[f(z)] = 0\quad\textrm{for }z\in(0,1)\,.
\ee
Since the real part of $f(z)$ vanishes for $z\in(0,1)$, we can easily cancel its branch cut in this region by writing
\be
f(z) = -\frac{2z-1}{[z(z-1)]^{1/2}}f_{1}(z)\,.
\ee
$f_1(z)$ is now analytic in the connected region $\mathbb{C}\backslash((-\infty,0]\cup[1,\infty))$ and satisfies\footnote{Note that the prefactor $(2z-1)/[z(z-1)]^{1/2}$ is symmetric under $z\mapsto1-z$, and not antisymmetric as one might think at first sight.}
\be
f_1(z) = f_1(1-z)\quad\textrm{and}\quad f_1(\bar{z}) = \overline{f_1(z)}\,.
\ee
In particular, $f_1(z)\in\mathbb{R}$ for $z\in(0,1)$. The asymptotic conditions on $f_1(z)$ are boundedness by $(z-1)^{-1/2+\epsilon}$ as $z\rightarrow 1$ and by $z^{-1-\epsilon}$ as $z\rightarrow\infty$. Note that
\be
\mathrm{Im}[f(z_b)] = \frac{2z_b-1}{[z_b(1-z_b)]^{1/2}}f_{1}(z_b)\,.
\ee
Thus, to satisfy positivity of $\mathrm{Im}[f(z_b)]$, we must have $f_1(z_b)>0$ if $z_b>1/2$ and $f_1(z_b)<0$ if $z_b<1/2$. Under this condition, our problem is equivalent to minimizing 
\be
\int_{1}^{\infty}\!\! dz\, \frac{2z-1}{[z(z-1)]^{1/2}}\frac{|f_{1}(z)|}{|f_1(z_b)|}\,.
\ee
There is a change of variables particularly convenient for our problem. It reads
\be
x(z) = \frac{\sqrt{z_b(1-z_b)}-\sqrt{z(1-z)}}{\sqrt{z_b(1-z_b)}+\sqrt{z(1-z)}}\,.
\ee
The map $z\mapsto x$ takes the region of analyticity of $f_1(z)$, i.e. $\mathbb{C}\backslash((-\infty,0]\cup[1,\infty))$, into the interior of the unit disk, and $z$ and $1-z$ are mapped to the same point. The point $z=z_b$ is mapped to $x=0$, and the points just above, below the branch cut $z\in(1,\infty)$ get mapped to the upper, lower half of the unit circle respectively. Defining $f_2(x)$ so that $f_2(x(z)) = f_1(z)$, we now want to minimize
\be
\oint\!\!\frac{dx}{2\pi i}\frac{1}{(x+1)^2}\frac{|f_{2}(x)|}{|f_2(0)|}\,,
\ee
where the contour of the integral is the unit circle. Here, we used the invariance of $dx/(1+x)^2$ under complex conjugation on the unit circle to double the contour from the upper semicircle to the whole unit circle. Finally, let us define
\be
f_3(x) = \frac{f_2(x)}{(1+x)^2}\,,
\ee
so that we want to minimize
\be
\frac{1}{2\pi}\!\int_{0}^{2\pi}\!\! d\theta\,\frac{|f_{3}\!\left(e^{i\theta}\right)\!|}{|f_3(0)|}\,.
\ee
Using Cauchy's theorem, we can write this as a ratio satisfying the inequality:
\be
\frac{\!\int_{0}^{2\pi}\!\! d\theta\,|f_{3}\!\left(e^{i\theta}\right)\!|}{\left|\int_{0}^{2\pi}\!\! d\theta\,f_{3}\!\left(e^{i\theta}\right)\right|}
\geq 1\,.
\ee
The inequality is only saturated if $f_3(x)$ is a constant. In order to prove that, first note that saturation requires $f_3(x)$ to have a constant phase on the whole unit circle. We can now define $f_3(x)$ outside the unit circle by
\be
f_3(x) = f_3(1/x)\,.
\ee
The constancy of the phase on the unit circle ensures $f_3(x)$ is continuous across the unit circle, and indeed holomorphic in the whole complex plane away from possible singularities at $x=\pm 1$. The asymptotic conditions on $f_1(z)$ at $z=1$ and $z=\infty$ imply that these singularities are removable. Moreover, $f_3(x)$ is bounded as $|x|\rightarrow\infty$ so it must in fact be a constant. This constant should be real and positive for $z_b>1/2$ and real and negative for $z_b<1/2$.

Tracing our way back, we can write the optimal $f(z)$ as
\be
f(z) = -\mathrm{sgn}(2z_b-1)\frac{2z-1}{[z(z-1)]^{1/2}\left[\!\sqrt{z(1-z)}+\sqrt{z_b(1-z_b)}\right]^2}\,,
\ee
which completes our solution to the original optimization problem.

\section{Fall-off conditions}
\label{app:falloff}
We will now review the conditions derived in \cite{Rychkov:2017tpc} for consistency of the class of functionals given in equation \reef{eq:ffg}. The functional action consists of two separate pieces involving the kernels $f(z)$, $g(z)$. We must both ensure that these pieces are {\em finite} and that the functional action commutes with crossing symmetry sum rules. The latter requirement can be stated as the {\em swapping} condition, according to which we must have for all $\Delta_\ell>0$
\bea
\lim_{\Delta^*\to \infty} \sum_{\Delta_\ell<\Delta<\Delta^*} c^2_{\Delta}\, \omega\left[F_{\Delta}^{\Df}\right]=\omega\left[\sum_{\Db<\Delta} F_{\Delta}^{\Df}\right],
\eea
or equivalently
\bea
\lim_{\Delta^*\to \infty}\omega\left[\sum_{\Delta^*<\Delta} F_{\Delta}^{\Df}\right]=0.
\eea
We will examine the implications of these conditions on the $f(z)$ and $g(z)$ pieces in turn.
It is clear that thanks to the  exponentially fast convergence of the OPE for any fixed $z$ in $\mathcal R$, the only way these conditions can fail is if the kernels do not decrease sufficiently fast near $z=1$ or $\infty$. 
We begin by noting that
\bea
F_{\De}^{\Df}(z)=O(|z|^{-2\Df}\log(|z|))\quad \textrm{as}\quad |z|\to \infty.
\eea
Hence finiteness of the $f(z)$ piece of the functional demands that $f(z)$ should fall off faster than $z^{2\Delta_\phi-1+\epsilon}$ at infinity, for some $\epsilon>0$. However the swapping property imposes a stronger condition. We have
\bea
\left|\sum_{\Delta^*<\Delta}c^2_\De F_{\Delta}^{\Df}(z)\right|
\leq \sum_{\Delta^*<\Delta}c^2_\Delta \left(\frac{|G_{\Delta}(\frac{z}{z-1})|}{|z|^{2\Delta_\phi}}+\frac{|G_{\Delta}(\frac{z-1}{z})|}{|1-z|^{2\Delta_\phi}}\right).
\eea
where we have used
\bea
|G_{\Delta}(\mbox{$\frac{z}{z-1}$})|=|G_{\Delta}(z)|\quad \textrm{for all}\quad z\in \mathbb C\backslash [1,+\infty). 
\eea
As $|z|\to \infty$, the arguments of the conformal blocks approach unity, and we probe the $u$-channel OPE limit. We may comfortably bound each sum on the RHS by the identity contribution in that channel, which grows like $|z|^{2\Df}$. Hence
\bea
\sum_{\Delta^*<\Delta} c_{\Delta}^2F_{\Delta}^{\Df}(z) =O(1)\quad \textrm{as}\quad z\to \infty.
\eea
We must therefore require the stronger fall off behaviour
\bea
f(z)\stackrel{|z|\to \infty}{\sim}o(z^{-1-\epsilon})\qquad \mbox{for some $\epsilon>0$}.
\eea
To conclude let us consider the contribution of $g(z)$. In this case the action of the functional is finite for an individual $F_\De^{\Df}$ if 
\bea
g(z)\stackrel{z\to 1}{\sim}o\left[(1-z)^{2\Df-1+\epsilon}\right]\qquad \mbox{for some $\epsilon>0$}.
\eea
Adding up infinite series of $F_{\Delta}^{\Df}$ cannot change this. This is because the only danger comes from the infinite sum of blocks in the direct channel, but this just reproduces the cross-channel singularities which are already taken into account by the condition above.

\small
\parskip=-10pt
\bibliography{Functionals}
\bibliographystyle{jhep}

\end{document}